\documentclass{article}

\usepackage{amssymb}

\usepackage{latexsym,amsmath,amsfonts,amscd}
\usepackage{xfrac}
\usepackage{graphics}
\usepackage{graphicx}
\usepackage{caption}
\usepackage{subcaption}
\usepackage{epstopdf}
\usepackage{color}
\usepackage{changebar}
\usepackage{indentfirst}
\usepackage{verbatim}
\usepackage{xcolor}
\usepackage{hyperref}
\usepackage{amssymb}
\usepackage{lscape}
\usepackage{array,multirow}

\usepackage[margin=3.0cm]{geometry}
\usepackage{authblk}
\usepackage{booktabs}

\renewcommand{\d}{{\rm d}}			
\DeclareMathOperator{\erf}{erf}
\newcommand{\CFL}{\mathsf{CFL}}		

\title{Uncertainty quantification of viscoelastic parameters in arterial hemodynamics with the a-FSI blood flow model}

\author[$\dagger$]{Giulia Bertaglia \footnote{Corresponding author. Email address: \textit{giulia.bertaglia@unife.it}}}
\author[$\star$]{Valerio Caleffi}
\author[$\dagger$]{Lorenzo Pareschi}
\author[$\star$]{Alessandro Valiani}

\affil[$\dagger$]{\small Department of Mathematics and Computer Science, University of Ferrara, Via Machiavelli 30, 44121 Ferrara, Italy}
\affil[$\star$]{\small Department of Engineering, University of Ferrara, Via G. Saragat 1, 44122 Ferrara, Italy}

\begin{document}

\maketitle

\begin{abstract}
This work aims at identifying and quantifying uncertainties related to elastic and viscoelastic parameters, which characterize the arterial wall behavior, in one-dimensional modeling of the human arterial hemodynamics. The chosen uncertain parameters are modeled as random Gaussian-distributed variables, making stochastic the system of governing equations. The proposed methodology is initially validated on a model equation, presenting a thorough convergence study which confirms the spectral accuracy of the stochastic collocation method and the second-order accuracy of the IMEX finite volume scheme chosen to solve the mathematical model. Then, univariate and multivariate uncertain quantification analyses are applied to the a-FSI blood flow model, concerning baseline and patient-specific single-artery test cases. A different sensitivity is depicted when comparing the variability of flow rate and velocity waveforms to the variability of pressure and area, the latter ones resulting much more sensitive to the parametric uncertainties underlying the mechanical characterization of vessel walls. Simulations performed considering both the simple elastic and the more realistic viscoelastic constitutive law show that the great uncertainty of the viscosity parameter plays a major role in the prediction of pressure waveforms, enlarging the confidence interval of this variable. In-vivo recorded patient-specific pressure data falls within the confidence interval of the output obtained with the proposed methodology and expectations of the computed pressures are comparable to the recorded waveforms.
\end{abstract}


\section{Introduction}
Over the last few decades, the availability of mathematical models of the human cardiovascular network has led to essential quantitative results in medical researches \cite{formaggia2009,quarteroni2004}. Numerical simulations provide efficient approaches for the quantification of hemodynamics variables, supplying meaningful data (not available by means of direct measurements or solely by invasive techniques) and even help the prediction of the possible onset of diseases and the evolution of pathologies \cite{toro2016a}. In recent years, mathematical models have been consistently developed, focusing on different aspects that need to be addressed to successfully model the circulatory system. Among these issues, the fluid-structure interaction (FSI) between blood flow and vessel wall gives rise to a complex viscoelastic mechanism that is difficult to describe properly with simple mathematical models and to efficiently simulate numerically \cite{valdez-jasso2009,westerhof2019}. An efficient way to characterize the viscoelastic FSI, including all its primary features (creep, stress relaxation and hysteresis), has been proposed in \cite{bertaglia2020}, presenting the augmented FSI (a-FSI) blood flow model together with a suitable IMEX finite volume discretization method. \par
The application of computational models to patient-specific simulations for clinical decision-making is one of the major challenges of recent years. Inevitably, the inputs to be personalized constitute possible sources of errors, given the large biological complexity and variability and because all measures are hampered by measurement uncertainty \cite{arnold2017,chen2013,eck2016}. Some of the uncertainties (i.e. epistemic uncertainties) can be reduced with more precise measurement or more advanced noise filtering techniques, while the others (i.e. random uncertainties) are very difficult if not impossible to capture accurately due to the inhomogeneous and multi-scale properties of the cardiovascular system which undergoes instantaneous changes in response to the heartbeat \cite{formaggia2009}. Sources of uncertainty which play an important role in the result of computational simulations can be divided into the following general categories \cite{arnold2017,chen2013,eck2016}:
\begin{enumerate}
\item \textit{Computational geometries}: Blood flow in the vascular system depends on vessel length, diameter and thickness of the wall (and, eventually, bends), whose uncertainty depends not only on the measurement technique adopted, usually recurring to magnetic resonance imaging (MRI) or computed tomography (CT), but also on the accuracy used during the measurement phase by the operator and the subject.
\item \textit{Fluid and vessel wall dynamics}: Blood flow can be associated to a Newtonian or a non-Newtonian rheology, and the choice of detailed or simplified velocity profiles affects the computation of friction losses. Moreover, the characterization of the wall mechanics through different mathematical models (rigid vessels, elastic or viscoelastic constitutive law) is linked to other sources of uncertainty.
\item \textit{Physical parameters}: Once the modeling is defined, model parameters such as blood viscosity, density, vessel stiffness, elasticity and viscoelastic relaxation are very difficult to measure, providing significant uncertainties even when single vessels are studied, as in this work.
\item \textit{Boundary conditions}: Uncertainties of boundary conditions, prescribed at the inlet and at the outlet of the computational domain, highly influence the confidence of the numerical solution. Inflow boundary conditions can be imposed through recorded inlet velocity or flow rate waveforms; while outflow boundary conditions are generally assigned recurring to the RCR or 3-element Windkessel model, which requires the definition of three additional parameters: two resistances and one compliance.
\item \textit{Numerical approximation}: In addition, errors associated to the numerical scheme used for the resolution of the mathematical model are another well-established issue.
\end{enumerate}
Hence, development and application of efficient computational methods for the assessment of the impact of parametric fluctuations on numerical solutions, referred as uncertainty quantification (UQ), is necessary for a correct interpretation of numerical simulations of arterial hemodynamics.\par
UQ methodologies can be classified into two classes of methods: \textit{intrusive} and \textit{non-intrusive}. Intrusive approaches can be very complex and hard to implement, since a parametrization of the uncertainty of the inputs is substituted into the model to derive new governing equations. Examples of these methods are the perturbation method, the momentum equation approach and the stochastic Galerkin method \cite{bijl2013,lemaitre2010,pareschi2020,pettersson2015,xiu2010}. In particular, stochastic Galerkin methods, based on generalized polynomial chaos (gPC) expansions, are very attractive thanks to the spectral convergence property with respect to the random input \cite{jin2017}. However, their intrusive nature leads to very challenging procedures for the reformulation of the problem, especially when the governing equations have complicated forms and when dealing with problems with multidimensional uncertainty (involving a large number of stochastic parameters). Intrusive approaches for systems of conservation laws with relaxation terms may cause the loss of important structural properties of the original problem, like hyperbolicity, well-balancing, positivity preserving and consistency of large time behavior \cite{jin2017,poette2009,xiu2010}.\par
On the other hand, non-intrusive methods do not require structural changes to deterministic numerical codes. Examples of non-intrusive methods are the Monte Carlo method and the stochastic collocation method \cite{eck2016,jin2017,xiu2010}. Monte Carlo sampling is extensively used, being a very robust and easy to parallelize approach since it only requires repetitive executions of deterministic simulations. Nevertheless, it can result unsuitable especially for systems that are already computationally expensive in their deterministic settings \cite{dimarco2014,jin2017}. Indeed, the solution statistics have a slow rate of convergence, with a mean value that converges as $\sfrac{1}{\sqrt{k}}$, with $k$ number of realizations \cite{xiu2010}. To address this issue, Multi-Level Monte Carlo approaches have been proposed recently, showing promising results for multivariate UQ \cite{mishra2012}, but loosing the ease of implementation. On the contrary, stochastic collocation method is a pseudo-spectral method that reflects the high accuracy of gPC methods, achieving exponential convergence rates when the solution is sufficiently smooth in the random space, with a straightforward implementation \cite{xiu2010,xiu2005}. The only issue related to this method is that computational costs increase very fast as the number of uncertain parameters increases, which is known as \textit{curse of dimensionality} \cite{xiu2010}. One of the major tools used to break the curse of dimensionality of grid-based approaches is the sparse grid technique, which permits a very efficient handling of stochastic simulations with a large number of random inputs \cite{jin2017,xiu2010,xiu2005}.\par
Several recent studies address the UQ problem in computational hemodynamics, investigating the effects of different sources of parametric uncertainty and adopting different stochastic frameworks. Some of them explore the sensitivity of blood flow and pressure to uncertainty in the inlet boundary condition \cite{arnold2017,brault2017} or in the outlet boundary condition \cite{chen2013,colebank2019}, with outcomes suggesting a high sensitivity on these uncertain inputs. More literature studies concern the impact of uncertainties related to geometrical and elastic mechanical parameters of vessels \cite{eck2017,petrella2019a,XIU2007}, but very few UQ analysis have been conducted considering the uncertainty underlying viscoelastic parameters that characterize the arterial wall behavior \cite{chen2013,perdikaris2014}.\par
The aim of this work is to investigate the effects of uncertainties of parameters involved in the elastic and viscoelastic constitutive equation on which it is based the recently proposed a-FSI blood flow model \cite{bertaglia2020,bertaglia2020a}. To this end, an IMEX finite volume stochastic collocation method is proposed, which combines:
\begin{itemize}
\item The stochastic collocation method, which guarantees spectral convergence in the stochastic space and ease of implementation compared to intrusive methods, avoiding the risk of loss of hyperbolicity of the approximated stochastic system of governing equations.
\item A second-order stiffly-accurate IMEX Runge-Kutta scheme that satisfies the AP property in the stiff limit (i.e. the scheme is consistent with the equilibrium limit, which corresponds to the asymptotic elastic behavior), hence for small relaxation times.
\item A second-order finite volume solver, which guarantees the correct treatment of the non-conservative terms of the hyperbolic model when computing fluxes and fluctuations.
\end{itemize}\par
The rest of the paper is structured as follows. In Section~\ref{section_a-FSIBloodFlow}, the a-FSI blood flow model is briefly presented, in its elastic and viscoelastic characterization. The numerical method is summarized in Section~\ref{sec:num_mod}, presenting the IMEX finite volume scheme, its AP property for the specific system of equations, and the stochastic collocation approach. In Section~\ref{section_UQBurgers}, a convergence study applied to the viscous Burgers equation (taken as model equation) is reported, for which the stochastic methodology is also presented in details. UQ analysis applied to the a-FSI blood flow model are presented and discussed in Section~\ref{section_UQBloodFlow}. Two baseline tests and two patient-specific tests are chosen, considering as vessels the upper thoracic aorta (TA), the common carotid artery (CCA) and the common femoral artery (CFA). For each test case, three univariate analysis are performed, followed by a multivariate one. Finally, conclusions are drawn in Section~\ref{section_conclusions}.
\section{The a-FSI blood flow model}
\label{section_a-FSIBloodFlow}
The standard 1D mathematical model for blood flow, valid for medium to large-size vessels, is obtained averaging the incompressible Navier-Stokes equations over the cross-section, under the assumption of axial symmetry of the vessel and of the flow, obtaining the well established equations of conservation of mass and momentum \cite{formaggia2009}. To close the resulting partial differential equations (PDE) system, a tube law, representative of the interaction between vessel wall displacement (through the cross-sectional area $A$) and internal blood pressure $p$, is required. In the simplest case, the pressure-area relationship is defined considering a perfectly elastic behavior of the vessel wall, with the widely adopted elastic constitutive tube law \cite{formaggia2003,muller2013,quarteroni2004}. But even though mathematical models representing the blood circulation frequently neglect the viscous component of the vessel wall, it is well known that blood vessels present viscoelastic properties \cite{nichols2011,salvi2012}. The Standard Linear Solid (SLS) model is yet able to exhibit all the three primary features of a viscoelastic material: creep, stress relaxation and hysteresis. Adding the constitutive equation derived from the SLS model in PDE form into the system of governing equations, leads to the following a-FSI system \cite{bertaglia2020,bertaglia2020a}:
\begin{subequations}
\begin{align}
	&\partial_t A + \partial_x (Au) = 0 \label{eq.cont}\\
	&\partial_t (Au) + \partial_x (Au^2)  + \frac{A}{\rho} \hspace{0.5mm} \partial_x p = f 		 \label{eq.mom}\\
	&\partial_t p + d \hspace{1mm}\partial_x (Au) = S \label{eq.PDE}.
\end{align}
\label{completesyst}
\end{subequations}
Here $u$ is the cross-section averaged blood velocity, $\rho$ is the blood density, $f$ is the friction loss term, $d$ is the parameter depending on the elastic component of the wall, while $S$ is the source term accounting for viscoelastic damping effects, and $x$ and $t$ are space and time respectively.\par
Blood velocity profile is considered self-similar and axisymmetric even in sections with large curvature (e.g. in the aortic arch). The choice of the typical velocity profile used for blood flow satisfying the no-slip condition \cite{alastruey2012} leads to a friction loss term which reads
\begin{equation}
f = -2(\zeta+2)\nu\pi u ,
\label{frictionterm}
\end{equation}
where $\nu$ is the kinematic viscosity of blood and $\zeta = \frac{2-\alpha_c}{\alpha_c-1}$ is a parameter depending on $\alpha_c$, the Coriolis coefficient. It has been demonstrated that the velocity profile is on average rather blunt in central arteries \cite{quarteroni2004}, with the consequence that the choice of $\alpha_c = 1.1$ ($\zeta = 9$) provides the best compromise to fit experimental data in these vessels. A parabolic velocity profile ($\alpha_c = \sfrac{3}{4}, \zeta = 2$) is instead more suitable for non-central arteries \cite{xiao2014}.\par
In Eq.~\eqref{eq.PDE}, the parameter $d$ represents the elastic contribution of the vessel wall:
\begin{equation}
d = \frac{K}{A} \left[m \left( \frac{A}{A_0}\right)^m - n \left( \frac{A}{A_0}\right)^n\right] ,
\label{d}
\end{equation}
where $A_0$ is the equilibrium cross-sectional area, $K$ represents the stiffness coefficient of the material and $m$ and $n$ are parameters associated to the specific behavior of the vessel wall, whether arterial or venous. In this work, dealing with arteries, the characterization of these parameters leads to:
\begin{equation}
K = \frac{E_0 h_0}{R_0}, \quad m = 1/2, \quad n = 0 ,
\label{K}
\end{equation}
with $E_0$ instantaneous Young modulus, $h_0$ wall thickness and $R_0$ equilibrium inner radius of the vessel. For further details, also regarding veins, the reader can refer to \cite{bertaglia2020,muller2013,toro2013}. The source term $S$ carries the viscous and damping information:
\begin{equation}
S  = \frac{1}{\tau_r}\left(\psi - p\right) .
\label{source}
\end{equation}
Here, $\tau_r$ is the relaxation time of the wall \cite{bertaglia2020}:
\begin{equation}
\tau_r = \frac{\eta \left(E_0 - E_{\infty}\right)}{E_0^2} ,
\label{relaxation_time}
\end{equation}
which depends on the three primary viscoelastic parameters of the SLS model: the instantaneous Young modulus $E_0$, the asymptotic Young modulus $E_{\infty}$ and the viscosity coefficient $\eta$ of the wall; while
\begin{equation}
\psi = K \frac{E_{\infty}}{E_0} \left[\left( \frac{A}{A_0}\right)^m - \left( \frac{A}{A_0}\right)^n\right] + p_{ext} ,
\label{psi}
\end{equation}
where $p_{ext}$ is the external pressure.\par
The reader is invited to notice how the formulation of the source term is coherent with the assumed mechanical behavior and consistent with the equilibrium limit. Indeed, when considering the elastic asymptotic limit, therefore $\tau_r \rightarrow 0$, even $S \rightarrow 0$. In fact, it can be observed from Eq.~\eqref{relaxation_time} that a vanishing relaxation time leads to a constant value of the Young modulus in time, with $\frac{E_{\infty}}{E_0} \rightarrow 1$, which exactly returns the limit of the classic elastic tube law \cite{bertaglia2020}.\par
Writing the non-linear non-conservative system \eqref{completesyst} in the general compact form leads to:
\begin{equation}
\partial_t \boldsymbol{Q} + \partial_x \boldsymbol{f}(\boldsymbol{Q}) + \boldsymbol{B}(\boldsymbol{Q}) \partial_x \boldsymbol{Q} = \boldsymbol{S}(\boldsymbol{Q}) ,
\label{systcompactform}
\end{equation}
in which
	\[\boldsymbol{Q} =
\begin{pmatrix} 
  	A \\ A u \\ p 
\end{pmatrix}, \quad
	\boldsymbol{f}(\boldsymbol{Q}) = 
\begin{pmatrix} 
 	A u \\ A u^2 \\ 0
\end{pmatrix}, \quad
	\boldsymbol{S}(\boldsymbol{Q}) = 
\begin{pmatrix} 
  	0 \\ f \\ S 
\end{pmatrix}, \quad
\boldsymbol{B}(\boldsymbol{Q}) =
\begin{pmatrix} 
  	0 &0 &0  \\ 0 &0 &\frac{A}{\rho} \\ 0 &d &0 
\end{pmatrix} .\]
Considering \(\boldsymbol{J}(\boldsymbol{Q}) = \partial \boldsymbol{f}/\partial \boldsymbol{Q} + \boldsymbol{B}(\boldsymbol{Q})\), it can also be expressed in the quasi-linear form:
\begin{equation}
\partial_t \boldsymbol{Q} + \boldsymbol{J}(\boldsymbol{Q})\partial_x \boldsymbol{Q} = \boldsymbol{S}(\boldsymbol{Q}) .
\label{systqlinearform}
\end{equation}
This system is hyperbolic, being the matrix \(\boldsymbol{J}(\boldsymbol{Q})\) diagonalizable, with a diagonal matrix containing all real eigenvalues, whose non-zero entries are $\lambda_1 = u-c , \lambda_2 = u+c$, and with a complete set of linearly independent eigenvectors \cite{bertaglia2020}. Here $c$ represents the wave speed:
\begin{equation}
c = \sqrt{\frac{A}{\rho} \hspace{0.5mm} d } = \sqrt{\frac{A}{\rho} \hspace{0.5mm} \frac{\partial p}{\partial A} } = \sqrt{\frac{K}{\rho} \left[m\left( \frac{A}{A_0}\right)^m - n\left( \frac{A}{A_0}\right)^n\right]}.
\label{eq:cel}
\end{equation} 
\subsection{Boundary conditions}
\label{section_BC}
The lumped-parameter model commonly used to simulate the effects of peripheral resistance and compliance on pulse wave propagation in large 1D arteries is the so-called RCR model (or 3-element Windkessel model), which consists of a resistor, with resistance $R_1$, connected in series with a parallel combination of a second resistor, with resistance $R_2$, and a capacitor, with compliance $C$ \cite{alastruey2008,xiao2014}.
At the outlet boundary of the 1D domain, the RCR model is coupled with the a-FSI blood flow model through the solution of the Riemann problem at the interface. Details of this coupling procedure are presented in \cite{bertaglia2020a}. Similarly, inflow boundary conditions are prescribed through an inlet flow rate or an inlet velocity (based on the available data), solving the Riemann problem at the inlet \cite{alastruey2012,bertaglia2020a}.
\section{Numerical method}
\label{sec:num_mod}
\subsection{IMEX Runge-Kutta finite volume scheme}
\label{section_IMEX}
Considering the $i$-$th$ cell $I_i=[x_{i+\frac{1}{2}},x_{i-\frac{1}{2}}]$ of a uniform mesh with grid size $\Delta x=x_{i+\frac{1}{2}}-x_{i-\frac{1}{2}}$, the discretization of system \eqref{systcompactform} by an IMEX Runge-Kutta finite volume scheme is summarized by:
\small
\begin{subequations}
\begin{align}
	&\boldsymbol{Q}^{(k)}_i = \boldsymbol{Q}^n_i -  \frac{\Delta t}{\Delta x} \sum_{j=1}^{k-1} \tilde{a}_{kj} \left[ \left( \boldsymbol{F}_{i + \frac{1}{2}}^{(j)}  - \boldsymbol{F}_{i - \frac{1}{2}}^{(j)} \right)  + \left(\boldsymbol{D}_{i+\frac{1}{2}}^{(j)}  + \boldsymbol{D}_{i-\frac{1}{2}}^{(j)}\right) + \boldsymbol{B}\left( \boldsymbol{Q}_i^{(j)}\right) \Delta \boldsymbol{Q}_i^{(j)} \right] + \Delta t \sum_{j=1}^{k} a_{kj} \boldsymbol{S}\left(\boldsymbol{Q}_i^{(j)}\right)
	\label{eq.iterIMEX} \\
	& \boldsymbol{Q}^{n+1}_i = \boldsymbol{Q}^n_i -  \frac{\Delta t}{\Delta x} \sum_{k=1}^{s} \tilde{b}_{k} \left[ \left( \boldsymbol{F}_{i + \frac{1}{2}}^{(k)}  - \boldsymbol{F}_{i - \frac{1}{2}}^{(k)} \right)  + \left(\boldsymbol{D}_{i+\frac{1}{2}}^{(k)}  + \boldsymbol{D}_{i-\frac{1}{2}}^{(k)}\right) + \boldsymbol{B}\left( \boldsymbol{Q}_i^{(k)}\right) \Delta \boldsymbol{Q}_i^{(k)} \right] + \Delta t \sum_{k=1}^{s} b_{k} \boldsymbol{S}\left(\boldsymbol{Q}_i^{(k)}\right)  \label{eq.finalIMEX}
\end{align}\label{eq.IMEX}
\end{subequations}
\normalsize
where $\Delta t = t^{n+1}-t^{n}$ is the time step satisfying the $\CFL$ condition and $\boldsymbol{Q}^n_i$ is the vector of the cell-averaged variables on $I_i$ at $t^{n}$.\par
An IMEX Runge-Kutta scheme is characterized by two $s \times s$ matrices, the explicit one, $\tilde a = (\tilde a_{kj})$, with $\tilde a_{kj} = 0$ for $ j\geq k$, and the implicit one, $a = (a_{kj})$, with $a_{kj} = 0$ for $j > k$, and by the weights vectors $\tilde b = (\tilde b_1, ...,\tilde b_s)^T$, $b = (b_1, ...,b_s)^T$ (with $s$ identifying the number of the Runge-Kutta stages). The distribution of these matrices leads to a scheme that treats implicitly the stiff terms (in the case here presented, the source terms) and explicitly all the non-stiff terms \cite{pareschi2005}. The explicit Runge-Kutta methods, indeed, are those for which non-zero entries in the $\tilde a$ matrix lie strictly below the diagonal. Entries at or above the diagonal will cause the right hand side of Eq.~\eqref{eq.iterIMEX} to involve $\boldsymbol Q_i^{(j)}$, giving a formally implicit method.\par
In the present work, the stiffly accurate IMEX-SSP2(3,3,2) Runge-Kutta scheme has been chosen, which is defined by the following explicit (on the left) and implicit (on the right) tableau in the usual Butcher notation \cite{pareschi2005}:
\begin{center}
\begin{tabular}{c | c c c}
0 & 0 & 0 & 0 \\
1/2 & 1/2 & 0 & 0 \\
1 & 1/2 & 1/2 & 0 \\ \hline
 & 1/3  & 1/3 & 1/3
\end{tabular}
\hspace{1.0cm}
\begin{tabular}{c | c c c}
1/4 & 1/4 & 0 & 0 \\
1/4 & 0 & 1/4 & 0 \\
1 & 1/3 & 1/3 & 1/3 \\ \hline
 & 1/3  & 1/3 & 1/3 
\end{tabular}
\end{center}
This scheme, characterized by $s=3$ stages for both the implicit part and the explicit part and 2-$nd$ order of accuracy, is AP and asymptotic accurate in the zero relaxation limit ($\tau_r \rightarrow 0$), which means that the consistency of the scheme with the equilibrium system is guaranteed and the order of accuracy is preserved in the stiff limit. To verify the AP property of the proposed scheme, let us concentrate on Eq.~\eqref{eq.PDE}, which depends on the relaxation time $\tau_r$, and express it following the IMEX Runge-Kutta discretization:
\begin{subequations}
\begin{align}
	&p = p^n e - \Delta t \tilde a \left(d\, \partial_x q\right) + \frac{\Delta t}{\tau_r} a \left(\psi - p\right) \label{eq.iterRK_p}\\
	&p^{n+1} = p^n - \Delta t \tilde b^T \left(d\, \partial_x q\right) + \frac{\Delta t}{\tau_r} b^T \left(\psi - p\right) \label{eq.finalRK_p} ,
\end{align}\label{eq.p}
\end{subequations}
where we denote $e^T = (1,\ldots,1) \in \mathbb{R}^s$. From Eq.~\eqref{eq.iterRK_p} we obtain
\begin{equation}
p = \left(\frac{\tau_r}{\Delta t} I + a\right)^{-1} \left(\frac{\tau_r}{\Delta t} p^n e -\tau_r\tilde a  \left(d\, \partial_x q\right) + a\psi \right),
\end{equation}
which leads to:
\begin{equation}
p = \frac{\tau_r}{\Delta t} a^{-1} p^n e - \tau_r a^{-1} \tilde a  \left(d\, \partial_x q\right) + \left( I - \frac{\tau_r}{\Delta t}a^{-1}\right) \psi + \mathcal{O}\left(\tau_r^2\right) .
\label{eq.iter_p_approx}
\end{equation}
If we now substitute Eq.~\eqref{eq.iter_p_approx} into Eq.~\eqref{eq.finalRK_p}, it results that:
\begin{equation}
\frac{\tau_r}{\Delta t} p^{n+1} = \frac{\tau_r}{\Delta t} \left(1 - b^T a^{-1}e\right)p^n + \tau_r \left(b^Ta^{-1} \tilde a - \tilde b^T\right)\left(d\, \partial_x q\right) + \frac{\tau_r}{\Delta t} b^T a^{-1} \psi,
\end{equation}
which implies consistency in the elastic (stiff) limit, for $\tau_r \rightarrow 0$. In fact, recalling from Eq.~\eqref{relaxation_time} that when $\tau_r \rightarrow 0$, then $E_{\infty} \rightarrow E_0$, in the elastic limit, from Eqs.~\eqref{eq.iter_p_approx} and \eqref{psi}, we recover exactly the elastic constitutive law \cite{bertaglia2020,muller2013}:
\begin{equation}
\label{eq.elastic_tube_law}
p = K \left[ \left( \frac{A}{A_0}\right)^m - \left( \frac{A}{A_0}\right)^n\right]+ p_{ext} .
\end{equation}
As emerged from this proof, another advantage of the chosen scheme lays in the possibility to obtain a totally explicit algorithm, avoiding the adoption of iterative procedures (e.g. Newton-Raphson method), with a consequent consistent reduction of the computational cost (the reader can refer to \cite{bertaglia2020} for further details).\par
To assure the correct treatment of both the conservative and the non-conservative part of system \eqref{systcompactform}, at each stage of the method, numerical fluxes and non-conservative jump terms are obtained applying the Dumbser-Osher-Toro (DOT) solver \cite{dumbser2011a,dumbser2011}:
\begin{equation}
\label{eq:flux}
	\boldsymbol{F}_{i\pm\frac{1}{2}} = \frac{1}{2} \left[ \boldsymbol{f}\left(\boldsymbol{Q}_{i\pm\frac{1}{2}}^{+}\right) + \boldsymbol{f}\left(\boldsymbol{Q}_{i\pm\frac{1}{2}}^{-}\right)\right] - \frac{1}{2} \int_{0}^{1} \left \lvert \boldsymbol{J}\left(\Psi\left(\boldsymbol{Q}_{i\pm\frac{1}{2}}^{-},\boldsymbol{Q}_{i\pm\frac{1}{2}}^{+},s\right)\right)\right \rvert \frac{\partial \Psi}{\partial s}\d s ,
\end{equation}
\begin{equation}
\label{eq:fluct}
	\boldsymbol{D}_{i\pm \frac{1}{2}} = \frac{1}{2} \int_{0}^{1} \boldsymbol{B}\left(\Psi\left(\boldsymbol{Q}_{i\pm \frac{1}{2}}^{-},\boldsymbol{Q}_{i\pm\frac{1}{2}}^{+},s\right)\right)\frac{\partial \Psi}{\partial s}\d s .
\end{equation}
The boundary-extrapolated values $\boldsymbol{Q}_{i\pm\frac{1}{2}}^{\pm}$ are evaluated using a total-variation-diminishing (TVD) approach, adopting the minmod slope limiter to achieve second-order accuracy also in space and to avoid spurious oscillations near discontinuities \cite{toro2009}.  Integrals in Eqs.~\eqref{eq:flux} and \eqref{eq:fluct} are approximated by a 3-points Gauss-Legendre quadrature formula after that a simple linear path $\Psi$, connecting left to right boundary values in the phase-space, has been chosen \cite{bertaglia2018}. More details concerning the implemented scheme can be found in \cite{bertaglia2020}.
\subsection{Stochastic Collocation Method}
\label{section_CM}
To present the stochastic model, we consider the probabilistic framework described in \cite{chen2013,xiu2005,XIU2007} and employ the concept of random variable in probability theory to randomize a deterministic parameter to a real-valued random variable $Y(\omega), \omega \in \Omega$, which is defined in a complete probability space $\left( \Omega,\mathcal{A},\mathcal{P}\right)$ consisting on a set of outcomes $\Omega$, the $\sigma$-algebra of events $\mathcal{A}$ and probability measure $\mathcal{P}$. Replacing the deterministic variable by the random one, the deterministic model of Eq.~\eqref{systqlinearform} becomes the following stochastic model:
\begin{equation}
\partial_t \boldsymbol{Q}(x,t,\omega) + \boldsymbol{J}(\boldsymbol{Q}(x,t,\omega))\partial_x \boldsymbol{Q}(x,t,\omega) = \boldsymbol{S}(\boldsymbol{Q}(x,t,\omega)) .
\label{syst_stochastic}
\end{equation}
We consider that the probability density function (PDF) $\rho_p:\Gamma \to \mathbb{R}^+$ of the random variable is known, in particular assuming (without loss of generality) that it has a Gaussian distribution with mean $\omega_m$ and standard deviation $\sigma$:
\begin{equation*}
\rho_{p}\left(\omega\right) = \frac{1}{\sqrt{2\,\pi\, \sigma^2}} e^{ -\frac{(\omega - \omega_{m})^2}{2\, \sigma^2} }
\end{equation*}
In this context, the stochastic state process depends on the uncertainties only through $Y$ in the image $\Gamma \equiv Y(\Omega)$ and the stochastic system can be seen as a parametrized system \cite{chen2013}. 

The solution of problem \eqref{syst_stochastic} can be computed employing a gPC expansion \cite{Ghanem1999,Ghanem1996}. In this approach, the approximated solution, $\tilde{\boldsymbol{Q}}^h(x,t,\omega)$, is expressed for each component of the vector as a finite series of orthonormal polynomials in terms of the stochastic variable:
\begin{equation}\label{eq:expansion}
\tilde{\boldsymbol{Q}}^h(x,t,\omega)=\sum_{j=1}^{M} \hat{\tilde{\boldsymbol{Q}}}_j(x,t) \phi_j(\omega),
\end{equation}
where $M$ is the number of terms of the truncated series and $\phi_j(\omega)$ are orthonormal polynomials, with respect to the measure $\rho_p(\omega)\, \d \omega$. For a given Gaussian distribution, polynomials $\phi_j(\omega)$ are chosen as Hermite polynomials \cite{Xiu2002}. 
Accordingly, the expansion coefficients $\hat{\tilde{\boldsymbol{Q}}}_j(x,t)$ can be conceptually obtained by:
\begin{equation}
\label{eq:exp_coeff_int}
\hat{\tilde{\boldsymbol{Q}}}_j(x,t) = \int_{\Gamma} \boldsymbol{Q}(x,t,\omega)\, \phi_j(\omega)\, \rho_p(\omega)\, \d \omega, \qquad j=1,\ldots,M.
\end{equation}\par
Following the stochastic collocation method \cite{Babuska2007}, the integrals for the expansion coefficients in Eq.~\eqref{eq:exp_coeff_int} are replaced by suitable quadrature $\mathcal{U}^{N_p}$ characterized by the set $\{\omega_m, w_m \}_{m=1}^{N_p}$, where $\omega_m$ is the $m$-th collocation point, $w_m$ is the corresponding weight and $N_p$ represents the number of quadrature points. For a random variable associated to a Gaussian PDF, a suitable quadrature is the Gauss-Hermite one, which reads:
\begin{equation}
\label{eq:exp_coeff_quad}
\hat{\boldsymbol{Q}}_j(x,t) = \mathcal{U}^{N_p}\left[\boldsymbol{Q}^d(x,t;\omega)\, \phi_j(\omega)\right] = \sum_{m=1}^{N_p}  \boldsymbol{Q}^d(x,t;\omega_m)\, \phi_j(\omega_m)\, w_m, \qquad j=1,\ldots,M
\end{equation}
where $\boldsymbol{Q}^d(x,t;\omega_m)$, with $m=1,\ldots,{N_p}$, is the deterministic solution of problem \eqref{systqlinearform} (or problem \eqref{syst_stochastic} for the fixed value $\omega_m$), obtained through the IMEX scheme presented in \S~\ref{section_IMEX}. In this way, the previously discussed AP property of the chosen IMEX scheme is preserved, leading to a stochastic asymptotic-preserving (sAP) scheme \cite{jin2015,jin2018}, which permits to switch from a stochastic collocation method for the viscoelastic problem to a stochastic collocation method for the elastic problem in a uniform way with respect to the involved parameters.\par
Finally, the complete approximated solution is given by:
\begin{equation}
\label{eq:expansion_final}
\boldsymbol{Q}^h(x,t,\omega)=\sum_{j=1}^{M} \hat{\boldsymbol{Q}}_j(x,t)\, \phi_j(\omega).
\end{equation}\par
After the computation of the expansion coefficients by Eq.~\eqref{eq:exp_coeff_quad} and their substitution in Eq.~\eqref{eq:expansion_final}, a convenient solution for further post-processing estimations is available. In particular, the computation of the random solution statistics can be easily evaluated. For instance, the expected (mean) value of $\boldsymbol{Q}(x,t,\omega)$,
\begin{equation*}
\label{eq:mean_an}
\mathbb{E}\left[\boldsymbol{Q}\right] =\int_{\Gamma} \boldsymbol{Q}(x,t,\omega)\, \rho_p(\omega)\, \d \omega ,
\end{equation*}
is approximated as:
\begin{equation}
\label{eq:mean_apx}
\mathbb{E}\left[\boldsymbol{Q}\right] \approx \mathbb{E}\left[\boldsymbol{Q}^h\right] =\int_{\Gamma} \boldsymbol{Q}^h(x,t,\omega)\, \rho_p(\omega)\, \d \omega \approx \sum_{m=1}^{N_p}  \boldsymbol{Q}^d(x,t;\omega_m)\, w_m
\end{equation}
if the same quadrature chosen for the approximation of the integral in Eq.~\eqref{eq:mean_apx} is used. Once the expectation is obtained, also the variance can be computed \cite{xiu2010}:
\begin{equation}
\label{eq:variance_apx}
\mathbb{V}\left[\boldsymbol{Q}\right] = \mathbb{E}\left[ \left( \boldsymbol{Q} - \mathbb{E}\left[\boldsymbol{Q}\right] \right)^2 \right] \approx \mathbb{E}\left[\left(\boldsymbol{Q}^h \right)^2\right] - \mathbb{E}\left[\boldsymbol{Q}^h \right]^2 .
\end{equation} \par
The advantages behind the choice of this stochastic collocation approach for our specific problem can be summarized as follows: 
\begin{itemize}
\item Stochastic collocation methods belong to the class of non-intrusive methods, hence they only require the evaluation of the solutions of the corresponding deterministic problems at each collocation point. Such a feature makes these methods very attractive especially for problems with complicated nonlinear governing equations like Eq.~\eqref{systcompactform} where the uncertainty is related to different elastic and viscoelastic parameters.
\item They avoid the loss of important structural properties of the original problem. For example, intrusive approaches applied to systems of balance laws, like the a-FSI blood flow model here treated, can lead to the loss of hyperbolicity \cite{poette2009}. 
\item They are pseudo-spectral methods that reflect the high accuracy of gPC approaches. When the solutions possess sufficient smoothness in the stochastic space, these methods have been proven to achieve an exponential convergence rate \cite{jin2017,xiu2010,xiu2005}.
\end{itemize}\par
A thorough convergence study of the method is presented also in this work in \S~\ref{section_resultsBurgers}, applied to a model equation. Further, the application study concerning UQ of arterial hemodynamics is fully discussed taking into account univariate and multivariate analysis. In the latter case, assuming the independence of the chosen stochastic variables, the joint PDF $\rho_p\left(\boldsymbol{\omega}\right)$ of the random vector is given by \cite{eck2016,xiu2010}:
\begin{equation*}
\rho_p\left(\boldsymbol{\omega}\right) = \prod_{k=1}^{N_s} \rho_{p,k}\left(\omega_k\right) ,
\end{equation*}
with $N_s$ number of stochastic parameters.
\section{UQ applied to a model equation}
\label{section_UQBurgers}
\subsection{The viscous Burgers equation}
\label{section_vBE}
The analysis of the proposed methodology is presented in a simplified framework, using an initial values problem (IVP) based on the viscous Burgers equation (vBE) as model problem \cite{petrella2019a}. The general IVP reads:
\begin{subequations}
\label{eq:IV_problem}
\begin{align}
&\partial_t q(x,t) + q(x,t) \partial_x q(x,t) = \nu \partial_{xx}^2 q(x,t); \label{eq:vBE_det} \\
&q(x,0) = q_0(x), \label{eq:vBE_IV}
\end{align}
\end{subequations}
where the field $q(x,t)$ depends on the space $x$ and the time $t$ and $\nu$ is the kinematic viscosity (or diffusion coefficient). The given function $q_0(x)$ defines the initial condition.\\
This problem can be solved applying the Cole–Hopf transformation \cite{Cole1951,Hopf1950}, with a reference solution that can be written explicitly, once initial conditions have been assigned. For the following analysis, a Gaussian initial condition is chosen and the details of the procedure for obtaining a quasi-exact solution of the problem are given in \ref{appendix:A}. Concerning numerical solutions, the application of the IMEX Runge-Kutta finite volume method presented in \S~\ref{section_IMEX} is discussed in \ref{appendix:B}.\par
Assuming $\nu$ as random variable, $q$ in Eq.~\eqref{eq:vBE_det} becomes a stochastic variable that is also function of the kinematic viscosity itself: $q(x,t,\nu)$. Thus, the deterministic IVP \eqref{eq:IV_problem} becomes the following stochastic IVP:
\begin{subequations}
\label{eq:IV_problem_Sto}
\begin{align}
&\partial_t q(x,t,\nu) + q(x,t,\nu) \partial_x q(x,t,\nu) = \nu \partial_{xx}^2 q(x,t,\nu) \label{eq:vBE_Sto} \\
&q(x,0,\nu) = q_0(x). \label{eq:vBE_IV_Sto}
\end{align}
\end{subequations}\par
The solution of the problem can be computed following the methodology presented in \S~\ref{section_CM}, solving the deterministic problem \eqref{eq:IV_problem} at each collocation point. In particular, it is underlined that, for the specific stochastic problem of the vBE here discussed, $q^d(x,t;\nu_m)$ can be computed using a quasi-exact approach (see \ref{appendix:A}) or numerically through the proposed IMEX Runge-Kutta scheme (see \ref{appendix:B}). \par
It is worth noting that, for the given problem and PDF associated to the random kinematic viscosity, a stochastic solution obtained with the procedure here presented is completely defined by the number of collocation points and by the kind of solution of the deterministic problem. For instance, a reference stochastic solution can be obtained using a quite large value for $N_p$ and the quasi-exact solution for the deterministic problem. Conversely, the ordinary application of the method prescribes the use of a reduced number of $N_p$ and the solution of the deterministic problem through the chosen numerical scheme.
\begin{table}[p!]
\centering
\caption{Error estimates and empirical order of accuracy of the expected value of the solution of the stochastic vBE, obtained applying the stochastic collocation method and computing the deterministic part using the quasi-exact solution reported in \ref{appendix:A}. $N_p$ indicates the number of points used for the stochastic collocation method.}
\label{tab:accuracy_mean_value_EX_SC_nGH}
\begin{tabular}{l c c c c c c}
\hline
$N_p$ & $L^1$ & $\mathcal{O}\left( L^1\right)$ & $L^2$ & $\mathcal{O}\left( L^2\right)$ & $L^{\infty}$ & $\mathcal{O}\left( L^{\infty}\right)$\\
\hline
      4 & 1.8618$\times 10^{-09}$ &         & 2.7279$\times 10^{-09}$ &         &  7.1109$\times 10^{-09}$ &         \\
      6 & 2.7363$\times 10^{-10}$ &  4.7291 & 4.9552$\times 10^{-10}$ &  4.2067 &  1.5424$\times 10^{-09}$ &  3.7692 \\
      8 & 4.2071$\times 10^{-11}$ &  6.5087 & 8.3597$\times 10^{-11}$ &  6.1860 &  2.9480$\times 10^{-10}$ &  5.7522 \\
     10 & 5.8690$\times 10^{-12}$ &  8.8270 & 1.3834$\times 10^{-11}$ &  8.0614 &  5.2417$\times 10^{-11}$ &  7.7397 \\
     12 & 9.6577$\times 10^{-13}$ &  9.8974 & 2.2977$\times 10^{-12}$ &  9.8467 &  1.0277$\times 10^{-11}$ &  8.9368 \\
     14 & 1.4371$\times 10^{-13}$ & 12.3587 & 3.8636$\times 10^{-13}$ & 11.5658 &  1.7218$\times 10^{-12}$ & 11.5892 \\
     16 & 2.3687$\times 10^{-14}$ & 13.5016 & 6.6042$\times 10^{-14}$ & 13.2290 &  3.3459$\times 10^{-13}$ & 12.2686 \\
\hline
\end{tabular}
\end{table}
\begin{table}[p!]
\centering
\caption{Error estimates and empirical order of accuracy of the variance of the solution of the stochastic vBE, obtained applying the stochastic collocation method and computing the deterministic part using the quasi-exact solution reported in \ref{appendix:A}. $N_p$ indicates the number of points used for the stochastic collocation method.}
\label{tab:accuracy_variance_EX_SC_nGH}
\begin{tabular}{l c c c c c c}
\hline
$N_p$ & $L^1$ & $\mathcal{O}\left( L^1\right)$ & $L^2$ & $\mathcal{O}\left( L^2\right)$ & $L^{\infty}$ & $\mathcal{O}\left( L^{\infty}\right)$\\
\hline
      4 & 2.5806$\times 10^{-11}$ &         & 3.3878$\times 10^{-11}$ &         &  6.5353$\times 10^{-11}$ &         \\
      6 & 2.8759$\times 10^{-12}$ &  5.4116 & 5.9102$\times 10^{-12}$ &  4.3063 &  2.2551$\times 10^{-11}$ &  2.6242 \\
      8 & 3.9799$\times 10^{-13}$ &  6.8746 & 1.0532$\times 10^{-12}$ &  5.9958 &  4.8976$\times 10^{-12}$ &  5.3081 \\
     10 & 4.9432$\times 10^{-14}$ &  9.3475 & 1.4211$\times 10^{-13}$ &  8.9762 &  5.7513$\times 10^{-13}$ &  9.5988 \\
     12 & 5.5714$\times 10^{-15}$ & 11.9731 & 1.7143$\times 10^{-14}$ & 11.6004 &  9.1420$\times 10^{-14}$ & 10.0873 \\
     14 & 6.1931$\times 10^{-16}$ & 14.2509 & 1.8864$\times 10^{-15}$ & 14.3166 &  8.5873$\times 10^{-15}$ & 15.3433 \\
     16 & 6.9080$\times 10^{-17}$ & 16.4256 & 2.1110$\times 10^{-16}$ & 16.4015 &  1.1512$\times 10^{-15}$ & 15.0485 \\
\hline
\end{tabular}
\end{table}
\begin{figure}[p!]
\centering
\includegraphics[width=0.8\textwidth]{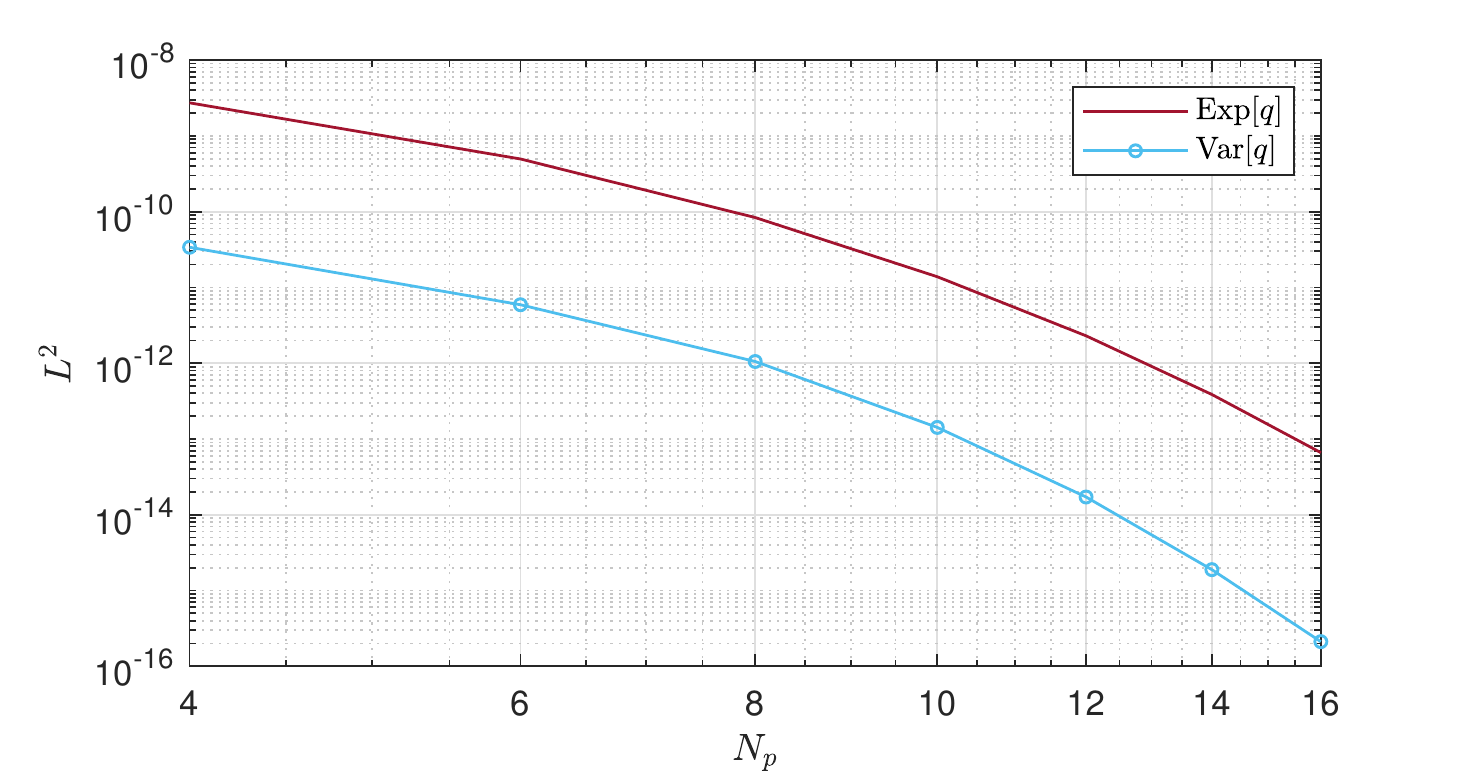}
\caption{Convergence study performed to test the accuracy of the stochastic collocation method applied to the vBE. Results shown in terms of $L^2$ norm errors of expected value and variance of the solution. $N_p$ indicates the number of points used in the stochastic collocation method.}
\label{fig:EX_SC_nGH}
\end{figure}
\begin{table}[t!]
\centering
\caption{Error estimates and empirical order of accuracy of the expected value of the solution of the stochastic vBE, with respect to the IMEX finite volume method and fixed grid of collocation points. $N_x$ indicates the number of cells in the physical domain.}
\label{tab:accuracy_mean_value_IMEX_SC_nc}
\begin{tabular}{l c c c c c c}
\hline
$N_x$ & $L^1$ & $\mathcal{O}\left( L^1\right)$ & $L^2$ & $\mathcal{O}\left( L^2\right)$ & $L^{\infty}$ & $\mathcal{O}\left( L^{\infty}\right)$\\
\hline
     99 & 1.1143$\times 10^{-2}$ &         & 5.6782$\times 10^{-3}$ &         &  5.7118$\times 10^{-3}$ &         \\
    297 & 1.6424$\times 10^{-3}$ &  1.7428 & 8.3112$\times 10^{-4}$ &  1.7491 &  7.3295$\times 10^{-4}$ &  1.8689 \\
    891 & 2.0144$\times 10^{-4}$ &  1.9101 & 1.0127$\times 10^{-4}$ &  1.9160 &  8.3604$\times 10^{-5}$ &  1.9761 \\
   2673 & 2.3091$\times 10^{-5}$ &  1.9716 & 1.1574$\times 10^{-5}$ &  1.9744 &  9.4153$\times 10^{-6}$ &  1.9877 \\
   8019 & 2.5934$\times 10^{-6}$ &  1.9902 & 1.2988$\times 10^{-6}$ &  1.9910 &  1.0434$\times 10^{-6}$ &  2.0024 \\
\hline
\end{tabular}
\end{table}
\begin{table}[t!]
\centering
\caption{Error estimates and empirical order of accuracy of the variance of the solution of the stochastic vBE, with respect to the IMEX finite volume method and fixed grid of collocation points. $N_x$ indicates the number of cells in the physical domain.}
\label{tab:accuracy_variance_IMEX_SC_nc}
\begin{tabular}{l c c c c c c}
\hline
$N_x$ & $L^1$ & $\mathcal{O}\left( L^1\right)$ & $L^2$ & $\mathcal{O}\left( L^2\right)$ & $L^{\infty}$ & $\mathcal{O}\left( L^{\infty}\right)$\\
\hline
     99 & 9.2671$\times 10^{-6}$ &         & 7.6418$\times 10^{-6}$ &         &  1.2095$\times 10^{-5}$ &         \\
    297 & 1.3025$\times 10^{-6}$ &  1.7861 & 1.0384$\times 10^{-6}$ &  1.8168 &  1.5558$\times 10^{-6}$ &  1.8667 \\
    891 & 1.5510$\times 10^{-7}$ &  1.9369 & 1.2142$\times 10^{-7}$ &  1.9535 &  2.0209$\times 10^{-7}$ &  1.8578 \\
   2673 & 1.7442$\times 10^{-8}$ &  1.9890 & 1.3531$\times 10^{-8}$ &  1.9973 &  2.0626$\times 10^{-8}$ &  2.0773 \\
   8019 & 1.9524$\times 10^{-9}$ &  1.9933 & 1.5135$\times 10^{-9}$ &  1.9939 &  2.4337$\times 10^{-9}$ &  1.9453 \\
\hline
\end{tabular}
\end{table}
\subsection{Convergence study}
\label{section_resultsBurgers}
To give a complete view of the proposed methodology, 3 accuracy analyses are performed. First of all, the exponential convergence of the stochastic collocation method described in \S~\ref{section_CM} is verified. Then, a convergence study of the method obtained coupling the stochastic collocation method with the IMEX Runge-Kutta scheme is performed both in the physical space and in the stochastic space.
All these analyses are performed considering the IVP given by Eq.~\eqref{eq:vBE_det} with the initial condition 
\begin{equation}
\label{eq:vBE_ic_gs}
q(x,0) = q_0(x) = q_{0,m} e^{-\frac{x^2}{2\sigma_m^2} },
\end{equation}
choosing the maximum value of the field $q_{0,m}=2.0$ and $\sigma_m=0.2$. The computational domain $[-L,L]$ is defined by $L=10$ and the final computational time is fixed at $t=3$.\par
To verify the spectral convergence of the stochastic method used, the kinematic viscosity is considered as random variable, with a Gaussian PDF having mean value $\nu_m=0.2$ and standard deviation $\sigma=0.01$.
The expected value of the solution and its variance are computed following the stochastic collocation approach. The deterministic part (i.e. the solution of the deterministic problem for a given value of $\nu$) is performed using the quasi-exact solution reported in \ref{appendix:A}. Numerical results are compared to reference solutions obtained using $N_p=100$ collocation points, in terms of expected value and variance. The expected exponential convergence is shown in Tables \ref{tab:accuracy_mean_value_EX_SC_nGH} and \ref{tab:accuracy_variance_EX_SC_nGH}, where $L^1$, $L^2$ and $L^{\infty}$ error norms and the related order of accuracy are presented. The result is highlighted by Fig.~\ref{fig:EX_SC_nGH}, in which the rapid decay of the $L^2$ norm error as the number of collocation points increases is observed in terms of expected value and variance.\par
In the convergence study of the method obtained coupling the stochastic collocation method with the IMEX Runge-Kutta scheme, the kinematic viscosity is again considered as a normally distributed stochastic variable. In the first case, to test the convergence in the physical space, the expected value of the solution and its variance are computed fixing $N_p=8$ collocation points. The reference solution is obtained using $N_p=100$ collocation points on each mesh, with the deterministic part computed following the quasi-exact approach (see \ref{appendix:A}). Tables \ref{tab:accuracy_mean_value_IMEX_SC_nc} and \ref{tab:accuracy_variance_IMEX_SC_nc} show the resulting error norms and confirm that the second-order of accuracy given in the physical space by the IMEX Runge-Kutta scheme is correctly achieved.\par
In the second case, to test the convergence in the stochastic space, expected value and variance of the solution are computed using the stochastic collocation method with different grids of collocation points. The deterministic solutions are performed using the IMEX Runge-Kutta scheme on a mesh with $N_x=891$ cells. For the solution over each grid, error norms are evaluated using the reference solution obtained considering $N_p=100$ collocation points on each mesh, with the deterministic part again computed through the quasi-exact approach. Clearly, in this case, the exponential convergence is not achieved (explicit results are omitted). To justify this behavior, in \cite{Xiu2007er} it has been shown that the error of the approximate gPC solution through the stochastic collocation scheme is given by the sum of three contributions. The first contribution is associated with the truncation errors of the gPC expansion \eqref{eq:expansion}, the second one is related to the error associated with the numerical solution of the IMEX Runge-Kutta scheme and the third one is related to the errors due to the quadratures of Eqs.~\eqref{eq:exp_coeff_quad} and \eqref{eq:mean_apx}. Given this, it is easy to observe that errors associated with the IMEX Runge-Kutta finite volume discretization (see Tabs.~\ref{tab:accuracy_mean_value_IMEX_SC_nc} and \ref{tab:accuracy_variance_IMEX_SC_nc}) are predominant compared to errors related to the stochastic collocation procedure (see Tabs.~\ref{tab:accuracy_mean_value_EX_SC_nGH} and \ref{tab:accuracy_variance_EX_SC_nGH}). Indeed, with the same grid refinement, the overall error in the computation of the expected value and the variance of the solution does not decay as the number of collocation points increases.
\begin{table}[t!]
\centering
\caption{Deterministic values of the model parameters considered for the viscoelastic baseline upper thoracic aorta (TA) and common carotid artery (CCA) and for the patient-specific common carotid artery (CCA-A) and common femoral artery (CFA-F), reported from \cite{bertaglia2020a}: vessel length $L$, inlet equilibrium radius $R_{0,in}$, outlet equilibrium radius $R_{0,out}$, vessel wall thickness $h_0$, reference celerity $c_0$, instantaneous Young modulus $E_0$, asymptotic Young modulus $E_{\infty}$, viscosity coefficient $\eta$, relaxation time $\tau_r$, external pressure $p_{ext}$, Coriolis coefficient $\alpha_c$, RCR model resistance $R_1$, RCR model resistance $R_2$, RCR model compliance $C$. In all the tests, $\rho = 1060$~kg/m$^3$, $\mu = 0.004$~Pa~s.}
\begin{tabular}{l c c c c }
\hline
	Parameter &TA &CCA &CCA-A &CFA-F\\
	\hline
	\(L\) $\left[\mathrm{cm}\right]$ &24.137	&12.60	&17.70 &14.50\\
	\(R_{0,in}\) $\left[\mathrm{mm}\right]$ &12.0	&3.0	&4.0	&3.7 \\
	\(R_{0,out}\) $\left[\mathrm{mm}\right]$ &12.0	&3.0	&3.7	&3.14\\
	\(h_0\) $\left[\mathrm{mm}\right]$ &1.2	&0.3	&0.3	&0.3 \\
	\(c_0\) $\left[\mathrm{m\hspace{0.5mm}s}^{-1}\right]$ &5.016 &6.635 &5.92 &7.05\\
	\(E_0\) $\left[\mathrm{MPa}\right]$ &0.7275	&1.7367	&1.7742	&2.2352 \\
	\(E_{\infty}\) $\left[\mathrm{MPa}\right]$ &0.5333	&0.9333	&0.9535	&1.2012 \\
	\(\eta\) $\left[\mathrm{kPa\hspace{0.5mm}s}\right]$ &23.884	&47.768	&47.768	&47.768 \\
	\(\tau_r\) $\left[\mathrm{s}\right]$ &0.009	&0.013	&0.0125	&0.010 \\
	\(p_{ext}\) $\left[\mathrm{mmHg}\right]$ &71.0		&82.0	&90.0	&90.0\\
	\(\alpha_c\) $\left[-\right]$ &1.1	&\sfrac{4}{3}	&\sfrac{4}{3}	&\sfrac{4}{3} \\
	\(R_{1}\) $\left[\mathrm{MPa\hspace{0.5mm}s\hspace{0.5mm}m}^{-3}\right]$ &11.752	&248.75	&145.91	&241.26\\
	\(R_{2}\) $\left[\mathrm{MPa\hspace{0.5mm}s\hspace{0.5mm}m}^{-3}\right]$ &111.67	&1869.7	&768.17	&2352.1\\
	\(C\) $\left[\mathrm{m}^{3} \mathrm{GPa}^{-1}\right]$ &10.163	&0.17529	&0.29178	&0.13208\\
\hline
\end{tabular}
\label{tab_testdata}
\end{table}
\begin{figure}[t!]
\begin{subfigure}{0.5\textwidth}
\centering
\includegraphics[width=1\linewidth]{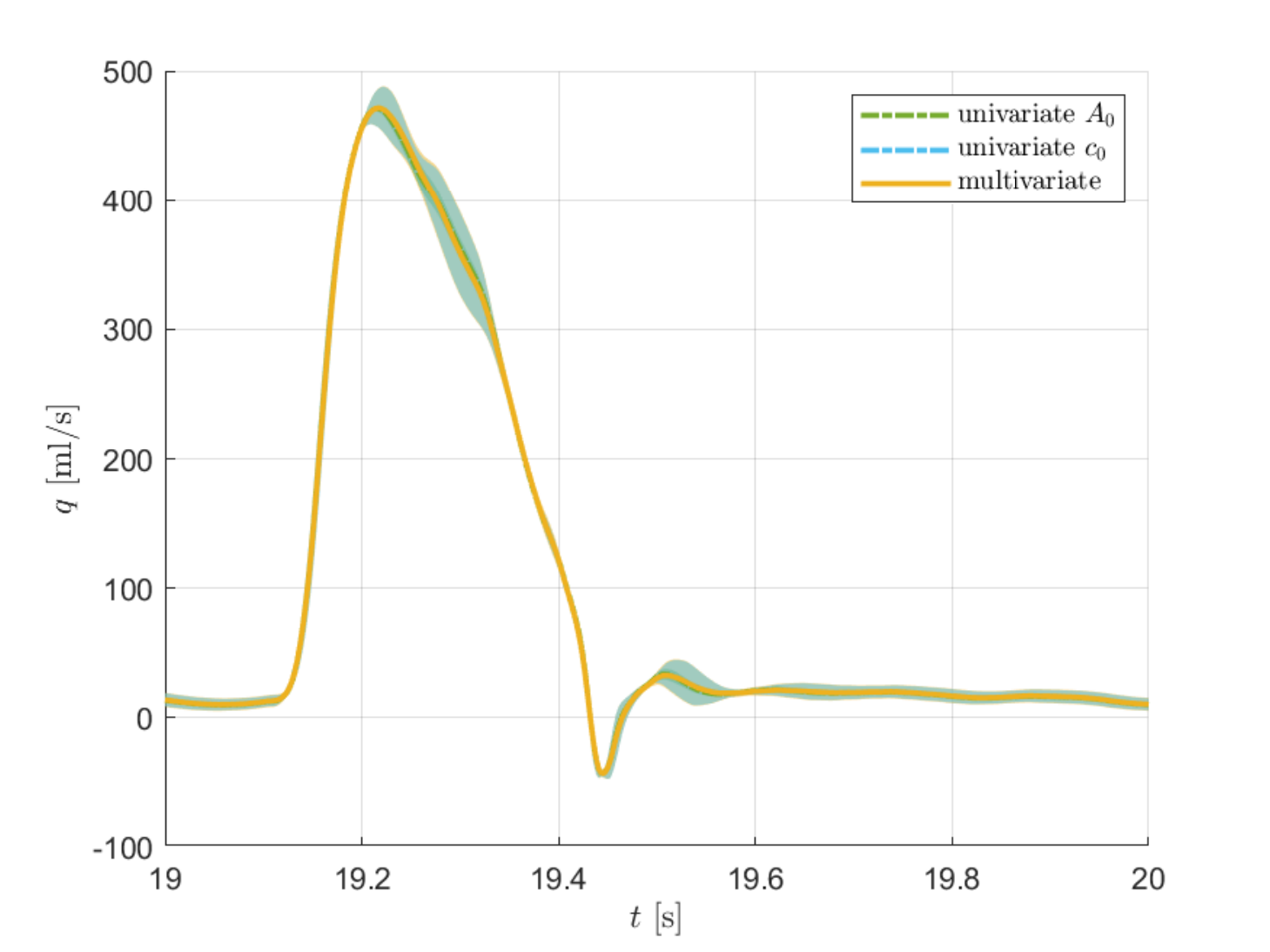}
\vspace*{-5mm}
\caption{}
\label{fig.TC14_elastic_q}
\end{subfigure}
\begin{subfigure}{0.5\textwidth}
\centering
\includegraphics[width=1\linewidth]{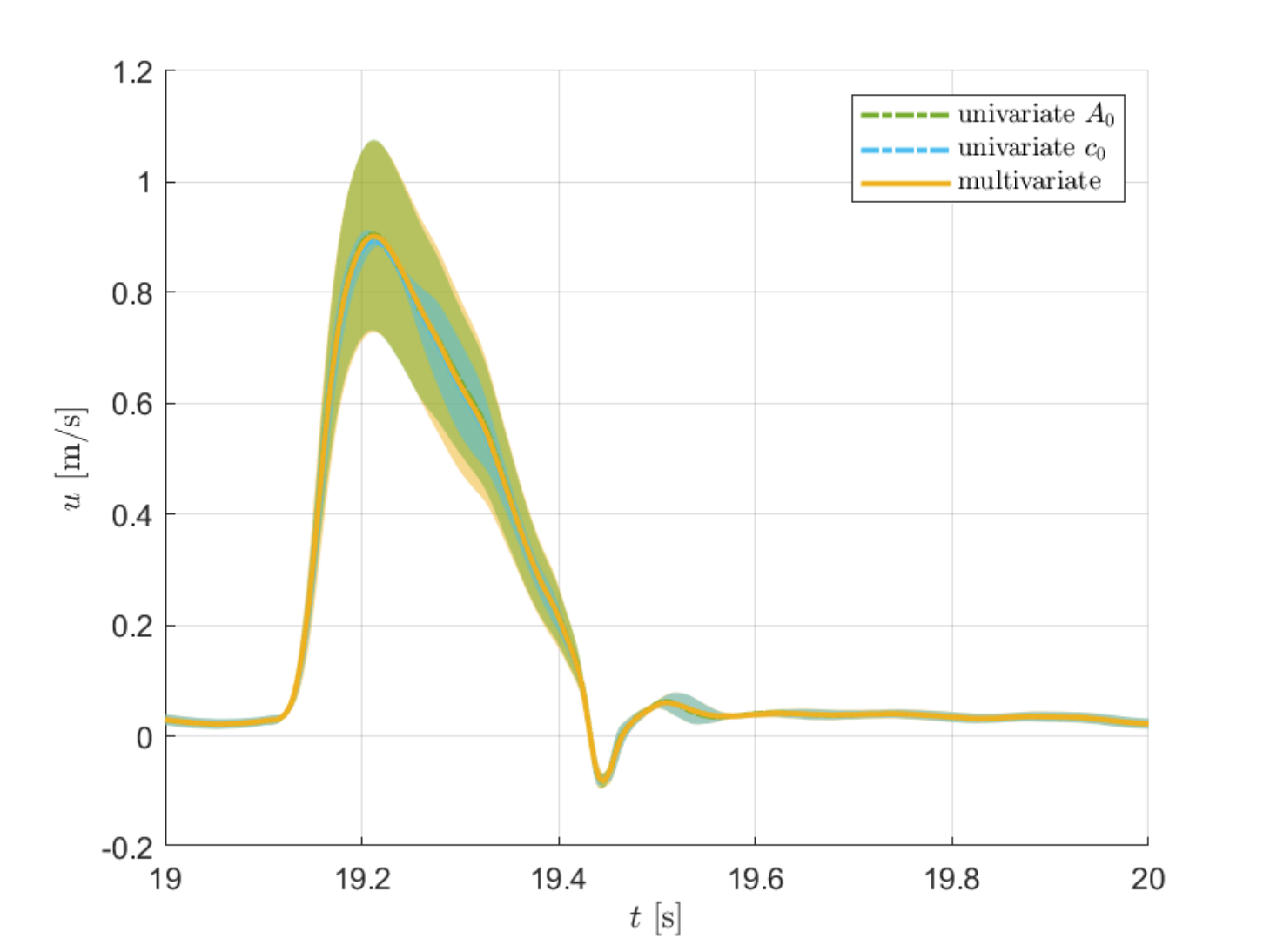}
\vspace*{-5mm}
\caption{}
\label{fig.TC14_elastic_u}
\end{subfigure}
\begin{subfigure}{0.5\textwidth}
\centering
\includegraphics[width=1\linewidth]{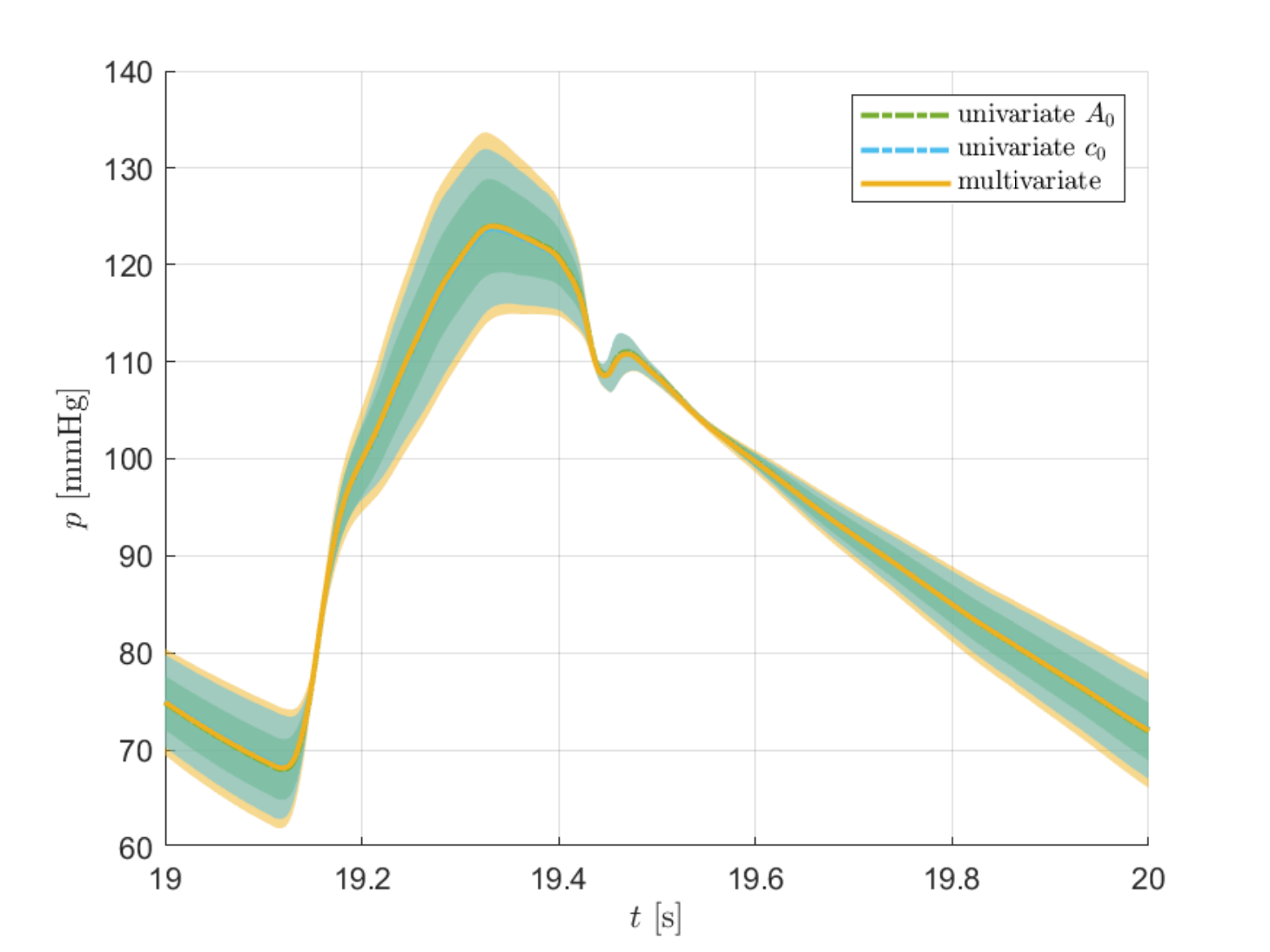}
\vspace*{-5mm}
\caption{}
\label{fig.TC14_elastic_p}
\end{subfigure}
\begin{subfigure}{0.5\textwidth}
\centering
\includegraphics[width=1\linewidth]{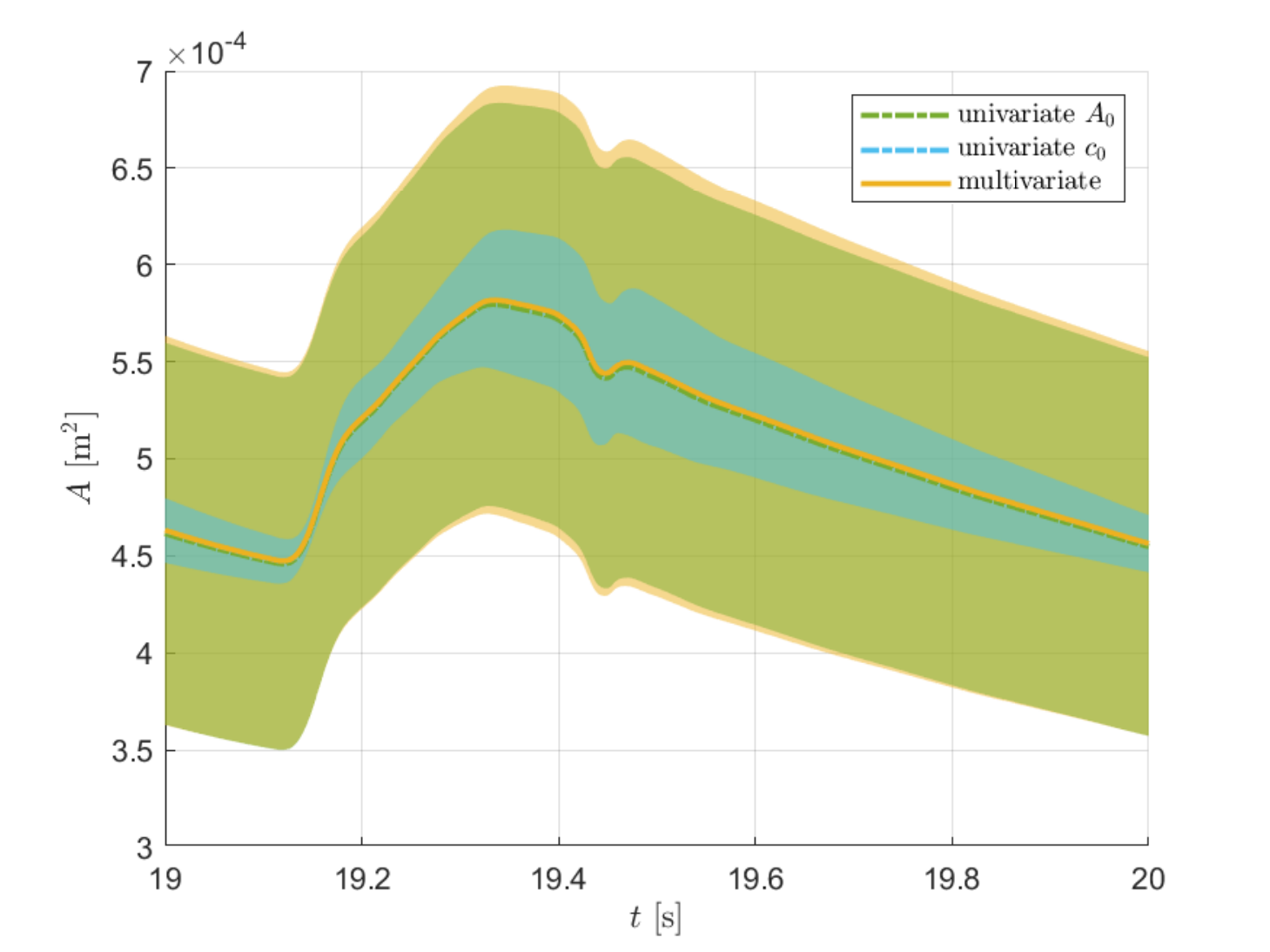}
\vspace*{-5mm}
\caption{}
\label{fig.TC14_elastic_A}
\end{subfigure}
\caption{Numerical results representative of one cardiac cycle obtained in the baseline TA test when characterizing the mechanical behavior of the vessel wall through an elastic law. Results are presented in terms of flow rate (a), velocity (b), pressure (c) and area (d) at the midpoint of the domain, for 95\% confidence intervals (colored area) and corresponding expectations (colored line). Each color is associated to a specific simulation, concerning the 2 univariate and the multivariate analysis.}
\label{fig.TC14_elastic}
\end{figure}
\begin{figure}[t!]
\begin{subfigure}{0.5\textwidth}
\centering
\includegraphics[width=1\linewidth]{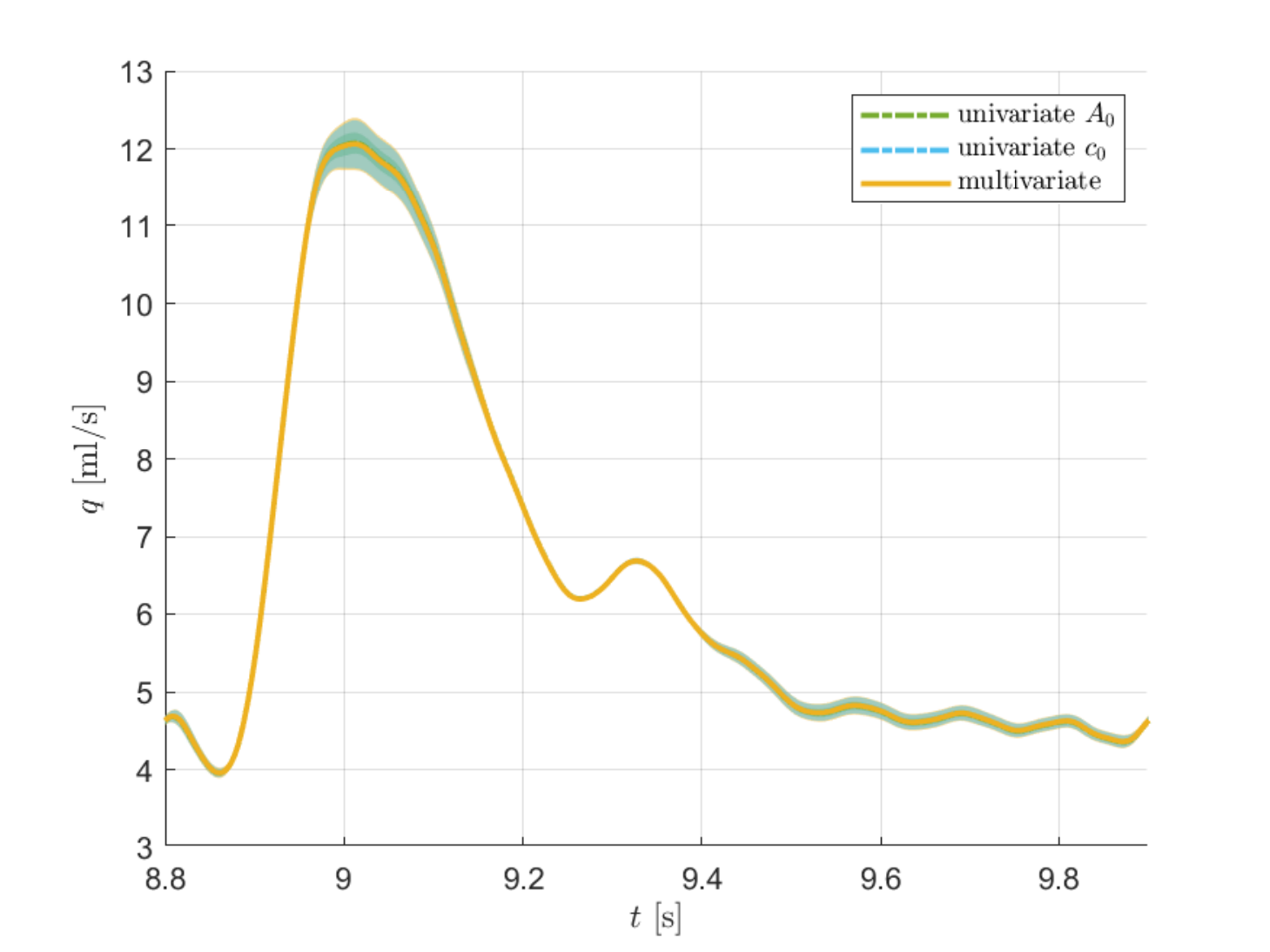}
\vspace*{-5mm}
\caption{}
\label{fig.TC15_elastic_q}
\end{subfigure}
\begin{subfigure}{0.5\textwidth}
\centering
\includegraphics[width=1\linewidth]{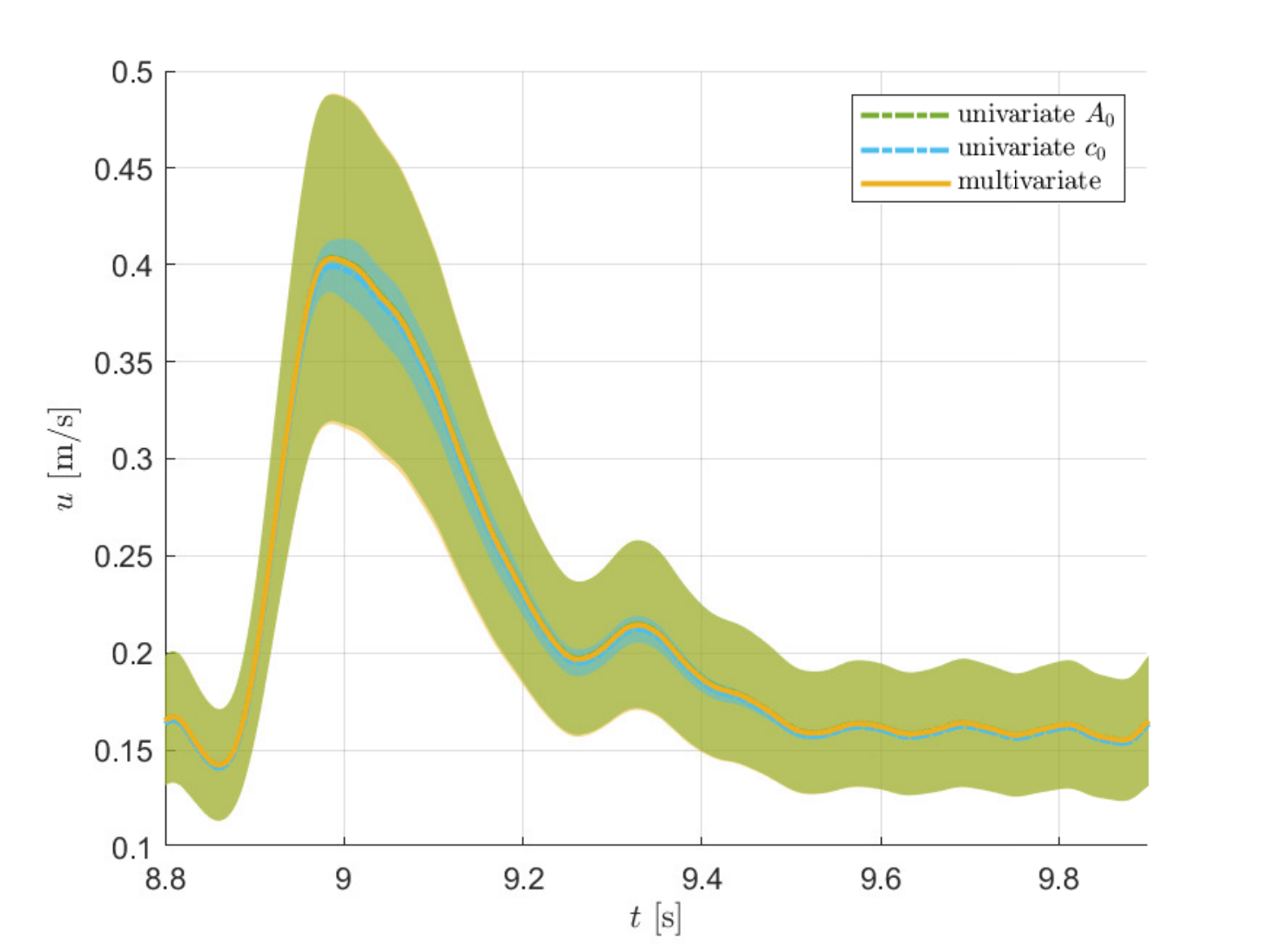}
\vspace*{-5mm}
\caption{}
\label{fig.TC15_elastic_u}
\end{subfigure}
\begin{subfigure}{0.5\textwidth}
\centering
\includegraphics[width=1\linewidth]{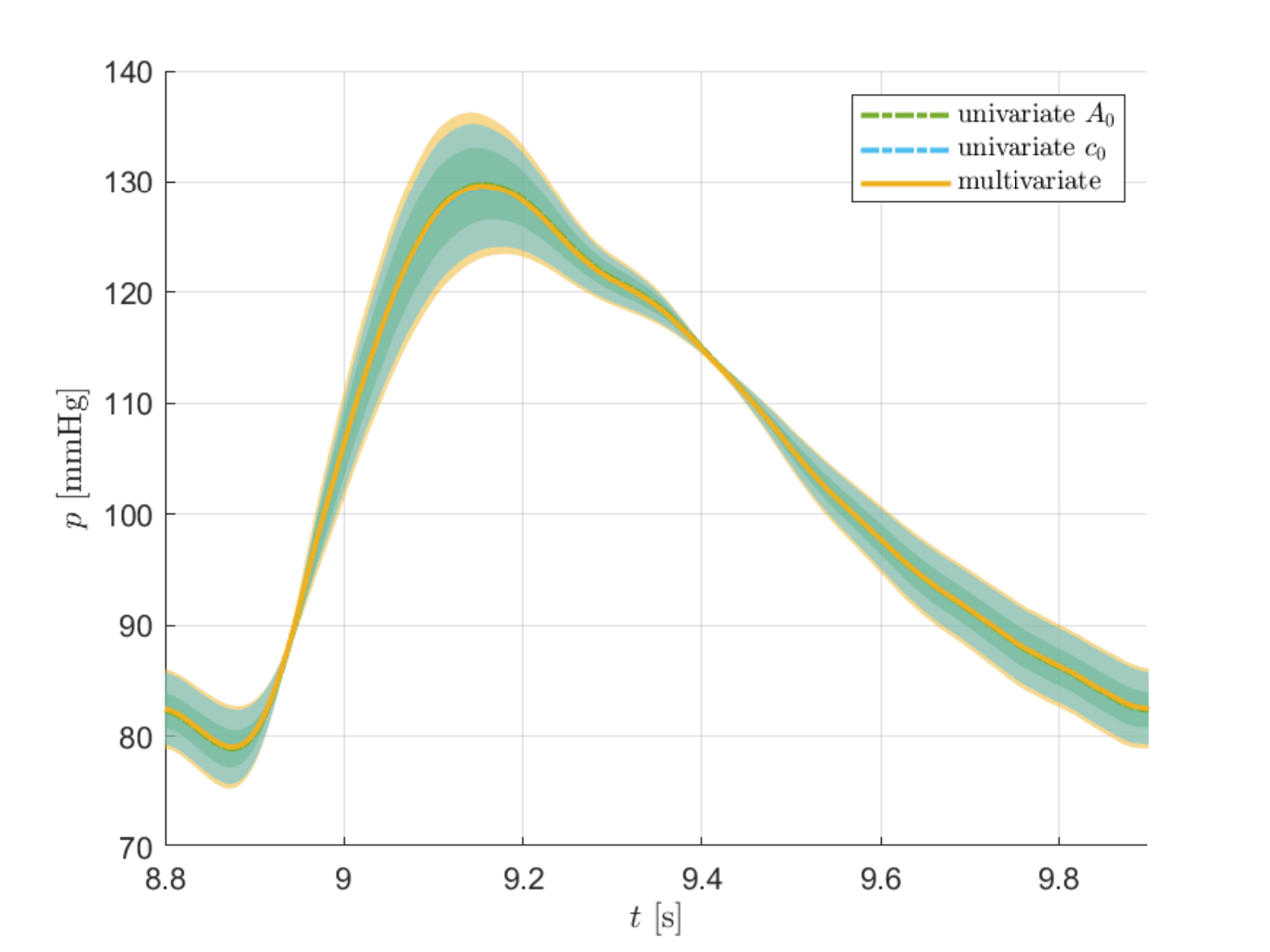}
\vspace*{-5mm}
\caption{}
\label{fig.TC15_elastic_p}
\end{subfigure}
\begin{subfigure}{0.5\textwidth}
\centering
\includegraphics[width=1\linewidth]{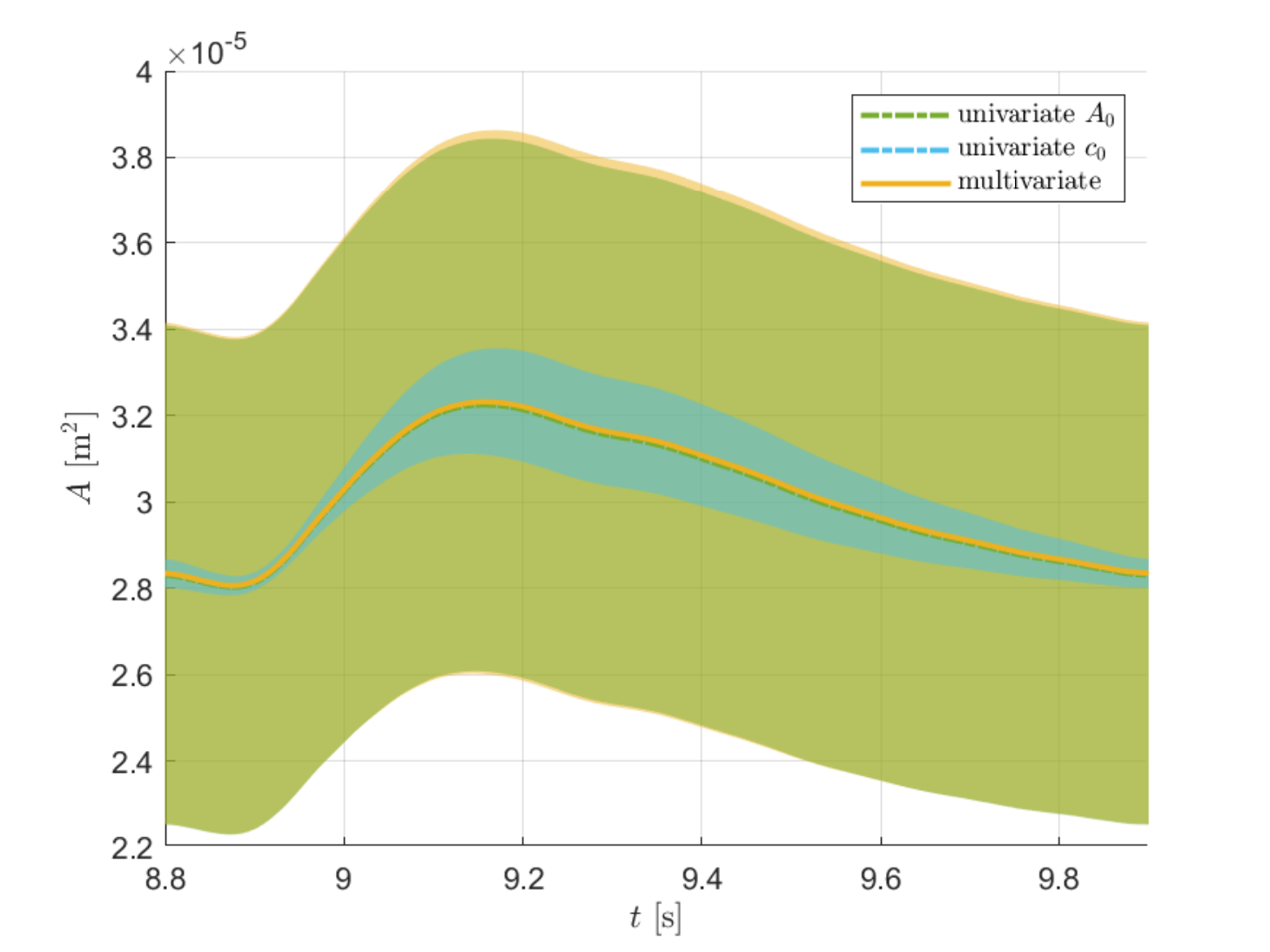}
\vspace*{-5mm}
\caption{}
\label{fig.TC15_elastic_A}
\end{subfigure}
\caption{Numerical results representative of one cardiac cycle obtained in the baseline CCA test when characterizing the mechanical behavior of the vessel wall through an elastic law. Results are presented in terms of flow rate (a), velocity (b), pressure (c) and area (d) at the midpoint of the domain, for 95\% confidence intervals (colored area) and corresponding expectations (colored line). Each color is associated to a specific simulation, concerning the 2 univariate and the multivariate analysis.}
\label{fig.TC15_elastic}
\end{figure}
\begin{figure}[t!]
\begin{subfigure}{0.5\textwidth}
\centering
\includegraphics[width=1\linewidth]{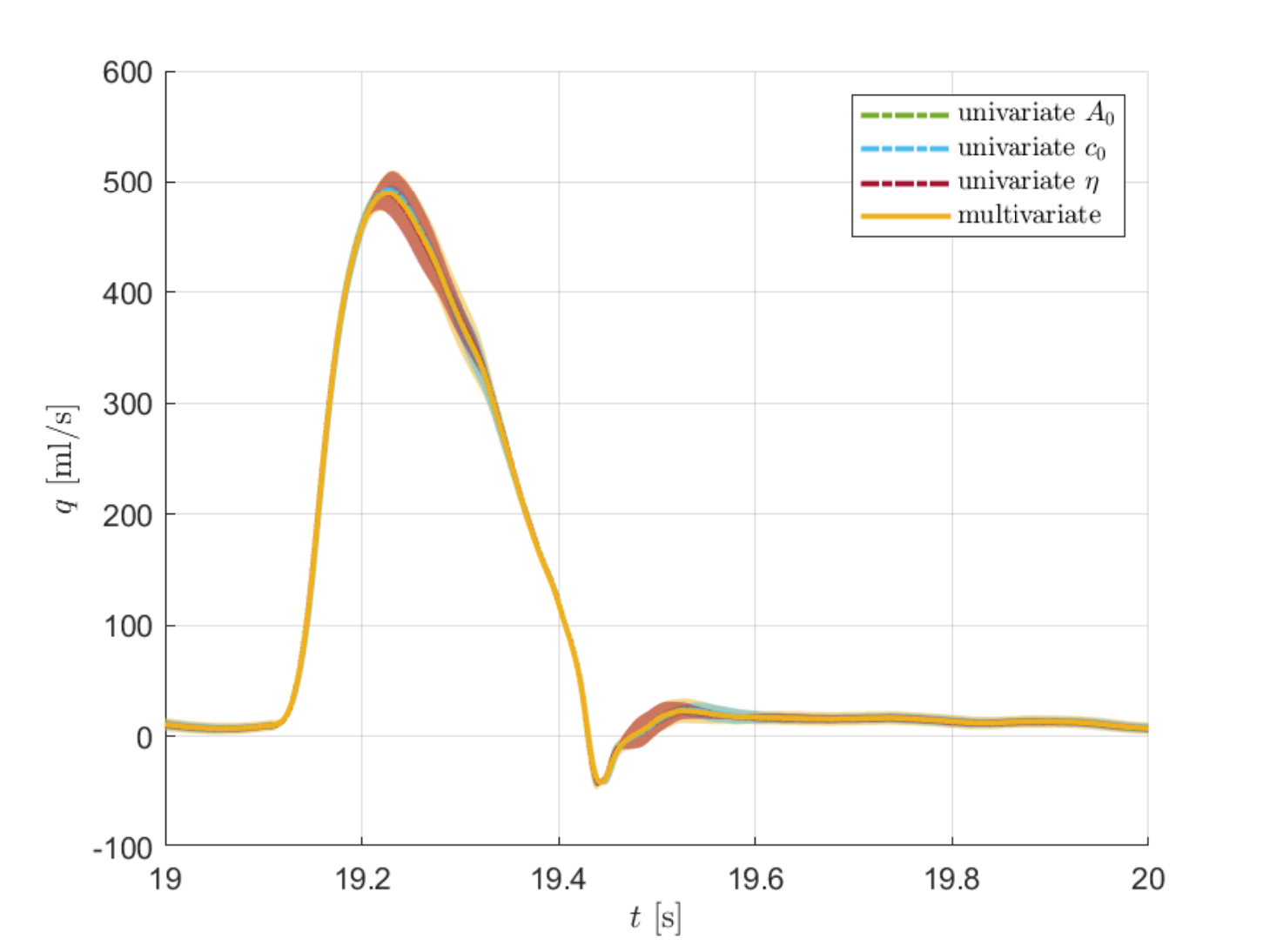}
\vspace*{-5mm}
\caption{}
\label{fig.TC14_visco_q}
\end{subfigure}
\begin{subfigure}{0.5\textwidth}
\centering
\includegraphics[width=1\linewidth]{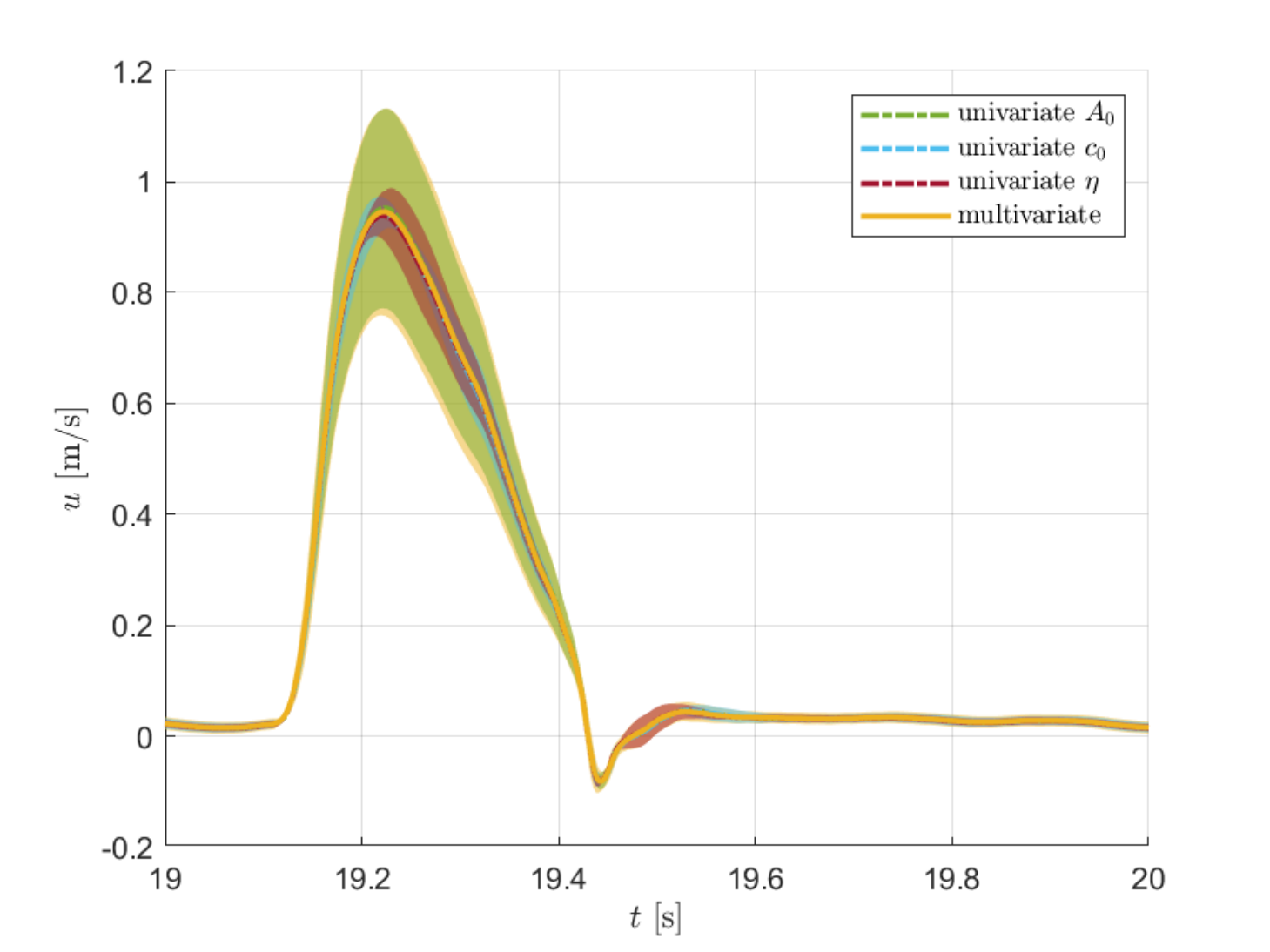}
\vspace*{-5mm}
\caption{}
\label{fig.TC14_visco_u}
\end{subfigure}
\begin{subfigure}{0.5\textwidth}
\centering
\includegraphics[width=1\linewidth]{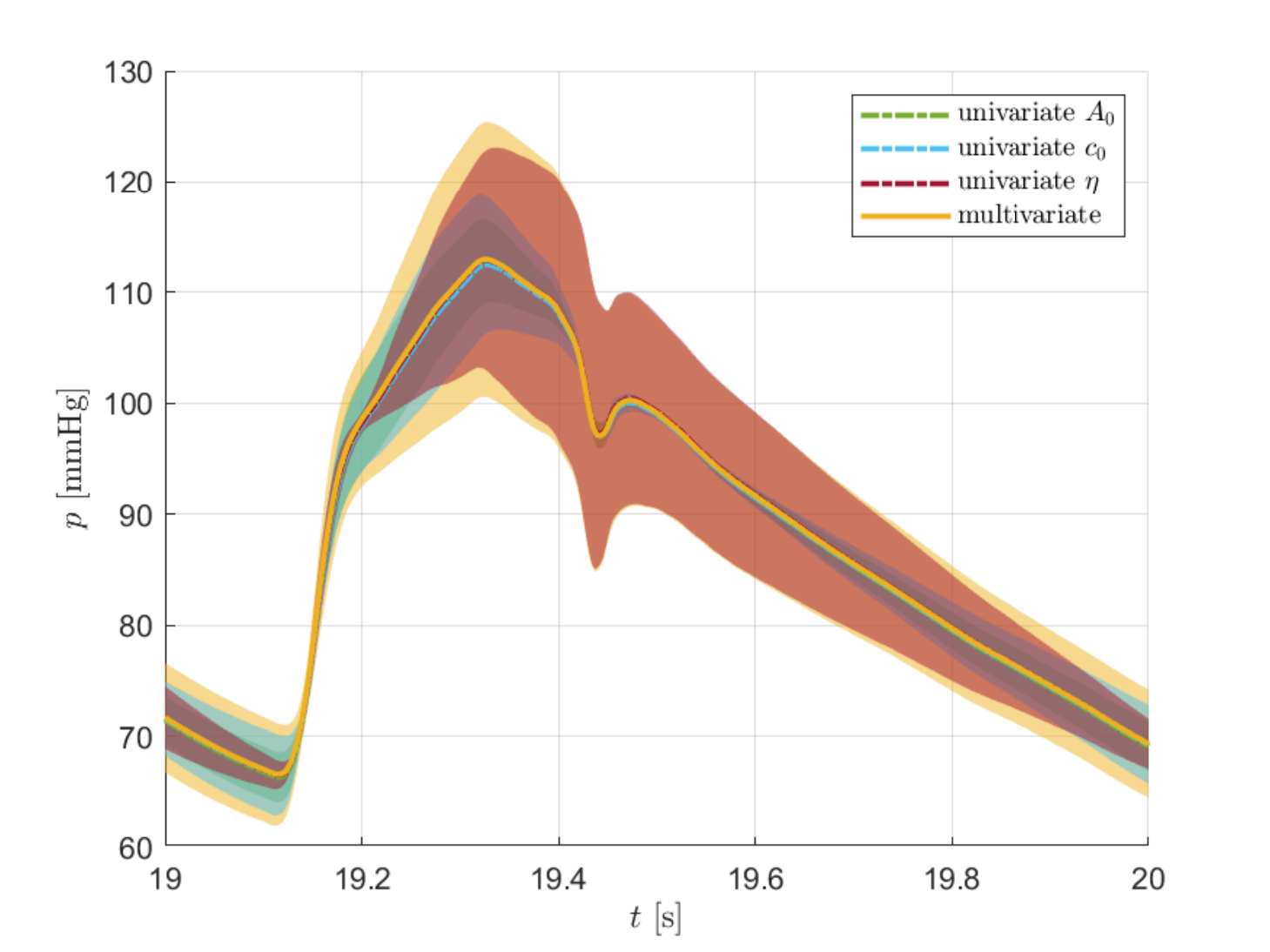}
\vspace*{-5mm}
\caption{}
\label{fig.TC14_visco_p}
\end{subfigure}
\begin{subfigure}{0.5\textwidth}
\centering
\includegraphics[width=1\linewidth]{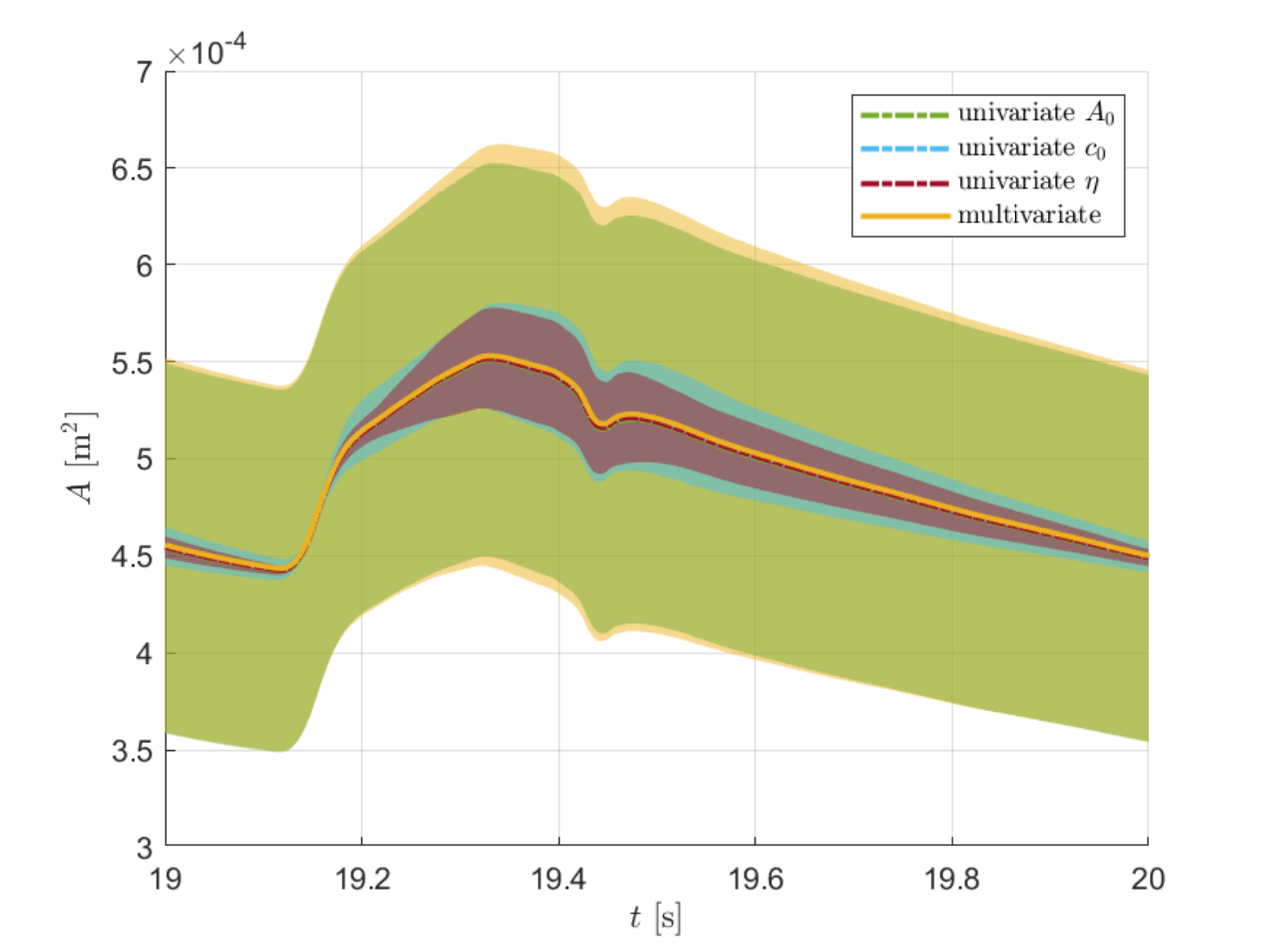}
\vspace*{-5mm}
\caption{}
\label{fig.TC14_visco_A}
\end{subfigure}
\caption{Numerical results representative of one cardiac cycle obtained in the baseline TA test when characterizing the mechanical behavior of the vessel wall through a viscoelastic law. Results are presented in terms of flow rate (a), velocity (b), pressure (c) and area (d) at the midpoint of the domain, for 95\% confidence intervals (colored area) and corresponding expectations (colored line). Each color is associated to a specific simulation, concerning the 3 univariate and the multivariate analysis.}
\label{fig.TC14_visco}
\end{figure}
\begin{figure}[t!]
\begin{subfigure}{0.5\textwidth}
\centering
\includegraphics[width=1\linewidth]{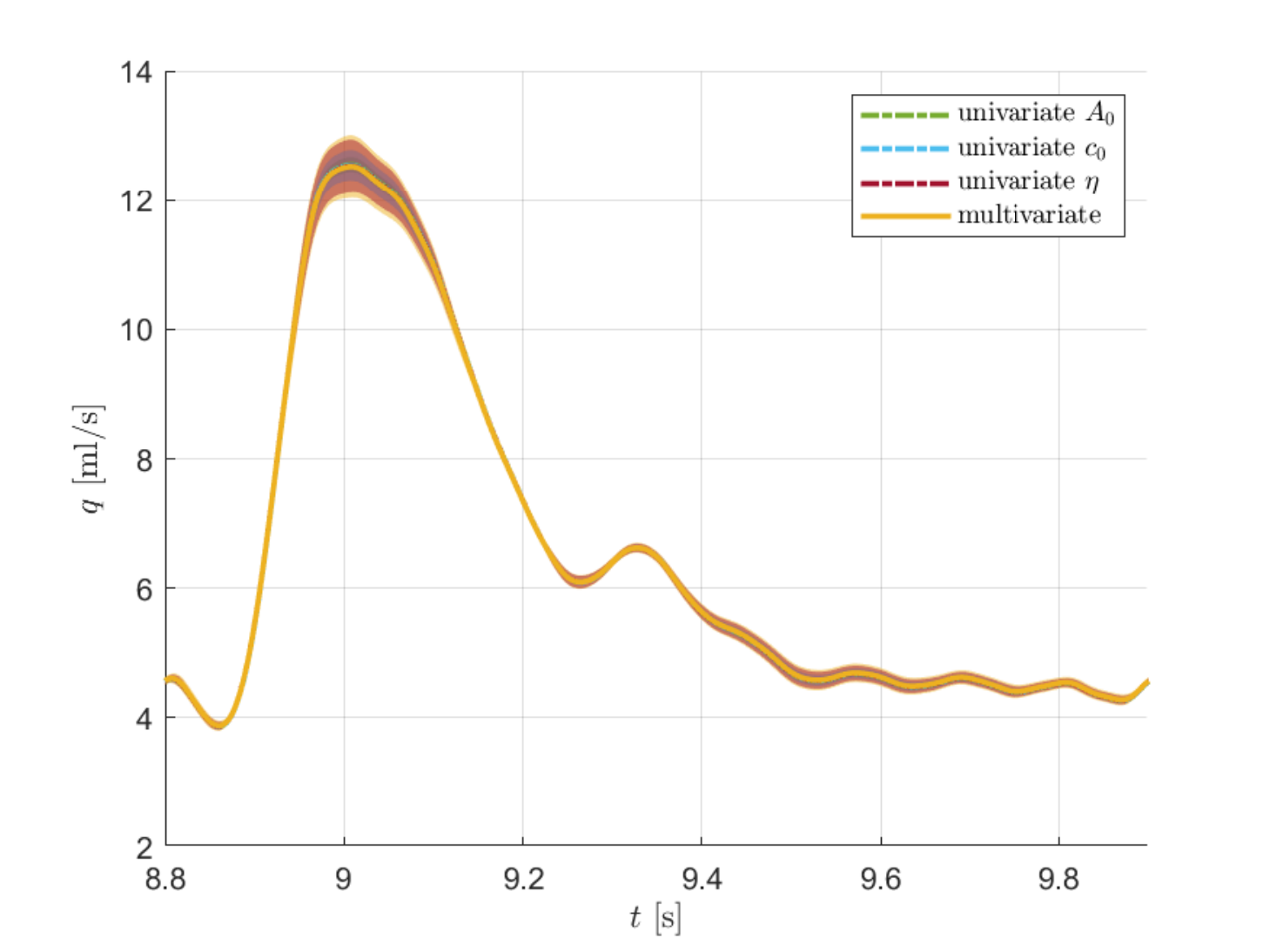}
\vspace*{-5mm}
\caption{}
\label{fig.TC15_visco_q}
\end{subfigure}
\begin{subfigure}{0.5\textwidth}
\centering
\includegraphics[width=1\linewidth]{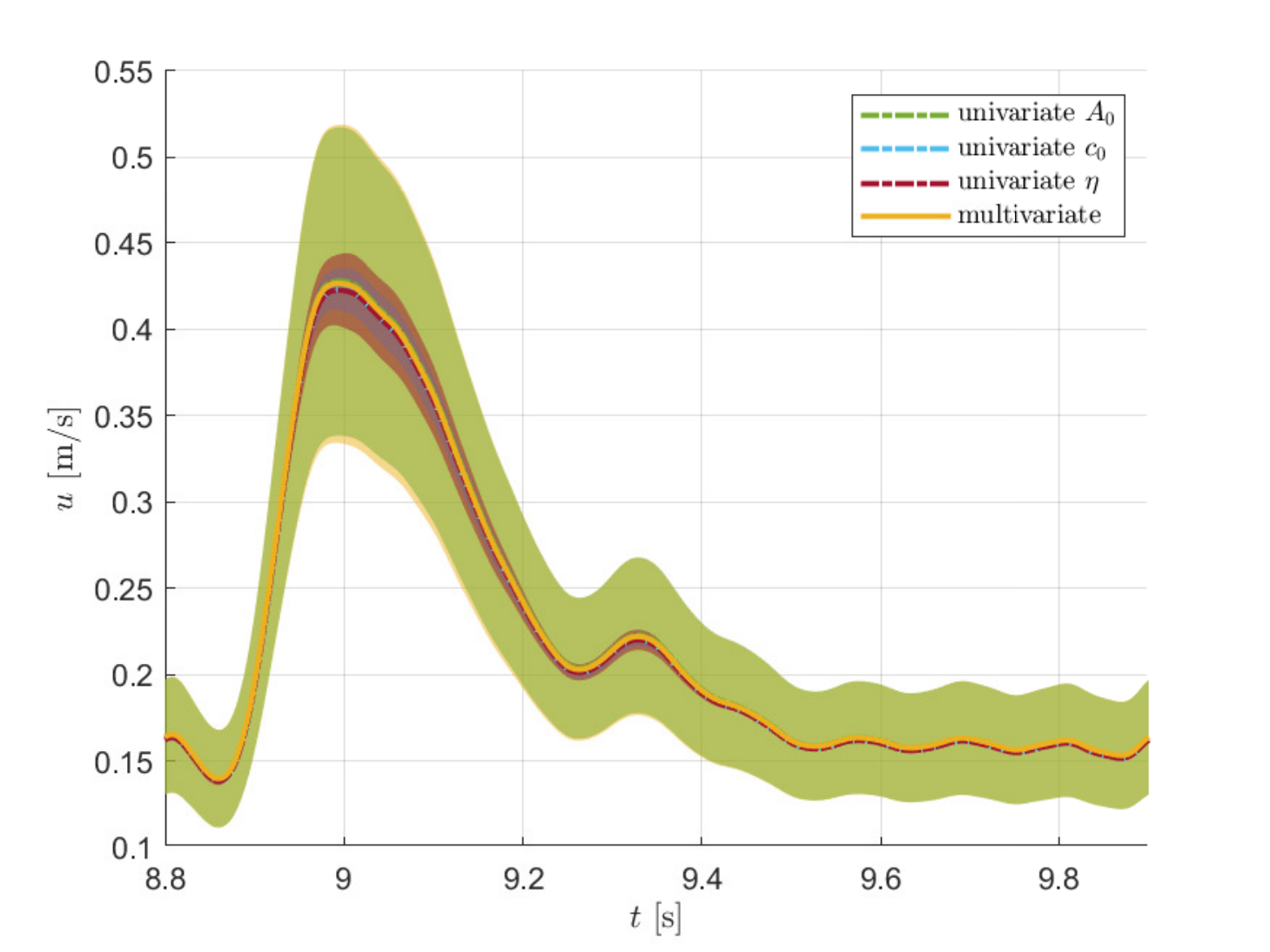}
\vspace*{-5mm}
\caption{}
\label{fig.TC15_visco_u}
\end{subfigure}
\begin{subfigure}{0.5\textwidth}
\centering
\includegraphics[width=1\linewidth]{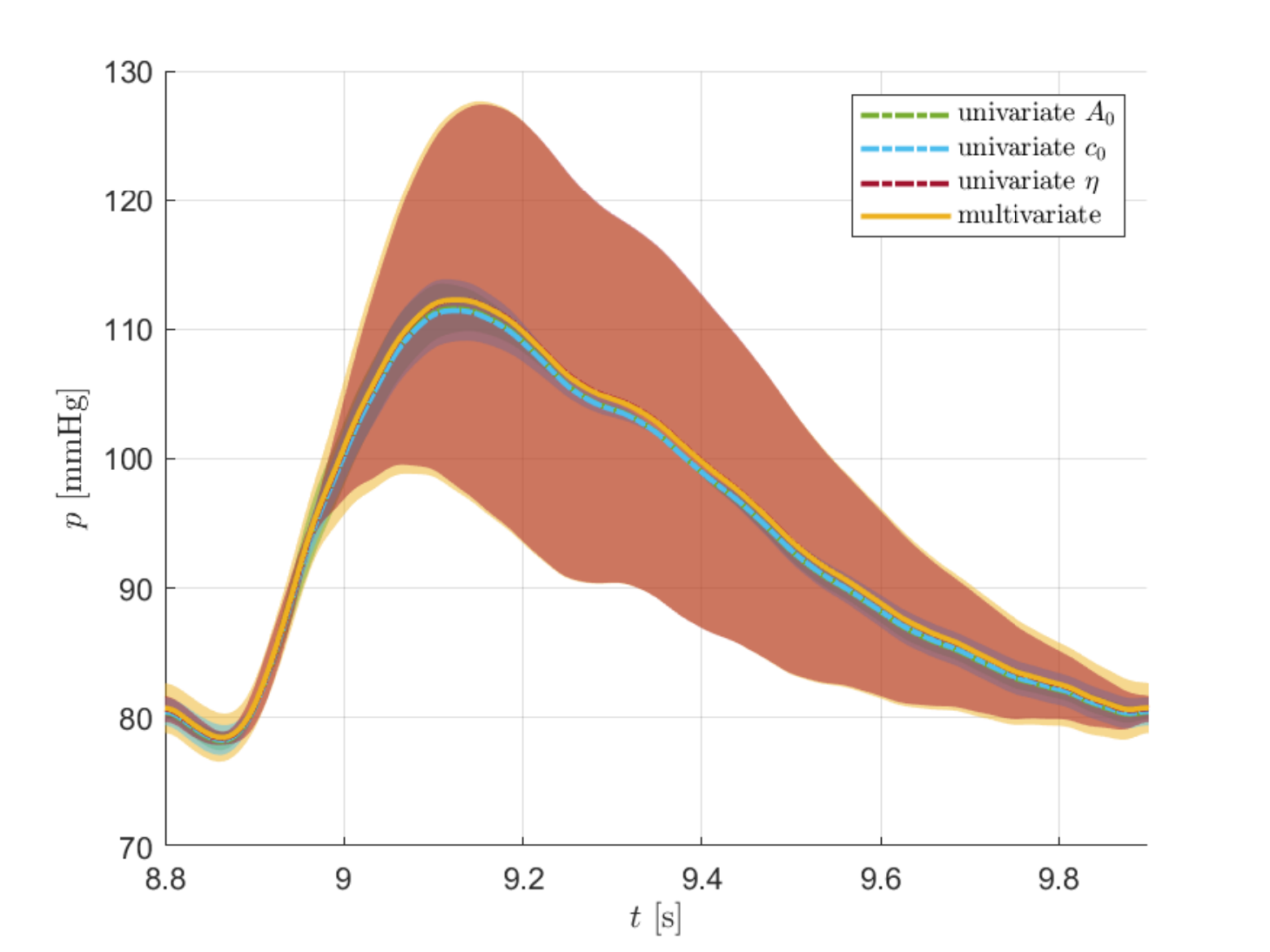}
\vspace*{-5mm}
\caption{}
\label{fig.TC15_visco_p}
\end{subfigure}
\begin{subfigure}{0.5\textwidth}
\centering
\includegraphics[width=1\linewidth]{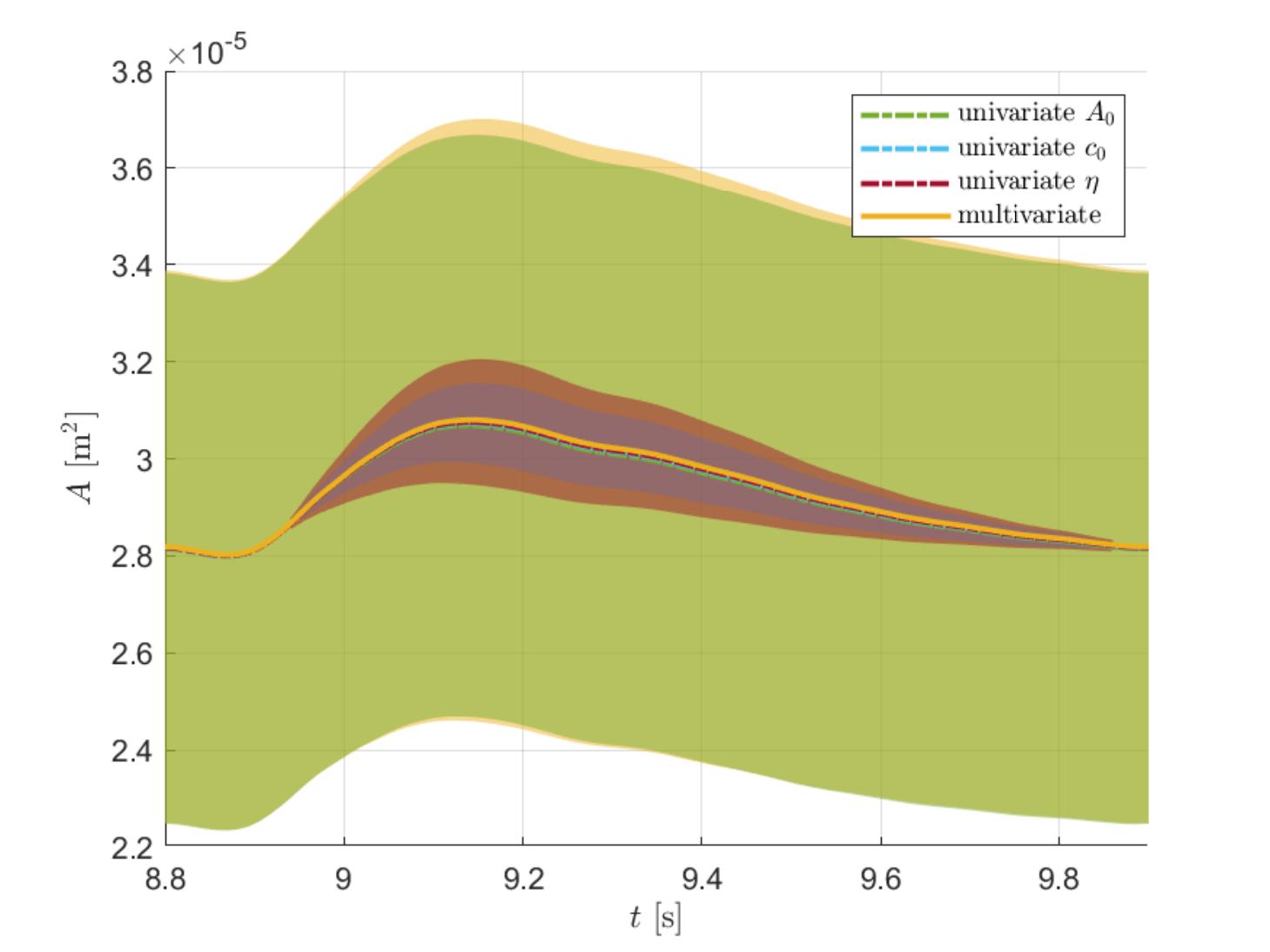}
\vspace*{-5mm}
\caption{}
\label{fig.TC15_visco_A}
\end{subfigure}
\caption{Numerical results representative of one cardiac cycle obtained in the baseline CCA test when characterizing the mechanical behavior of the vessel wall through a viscoelastic law. Results are presented in terms of flow rate (a), velocity (b), pressure (c) and area (d) at the midpoint of the domain, for 95\% confidence intervals (colored area) and corresponding expectations (colored line). Each color is associated to a specific simulation, concerning the 3 univariate and the multivariate analysis.}
\label{fig.TC15_visco}
\end{figure}
\begin{figure}[t!]
\begin{subfigure}{0.5\textwidth}
\centering
\includegraphics[width=1\linewidth]{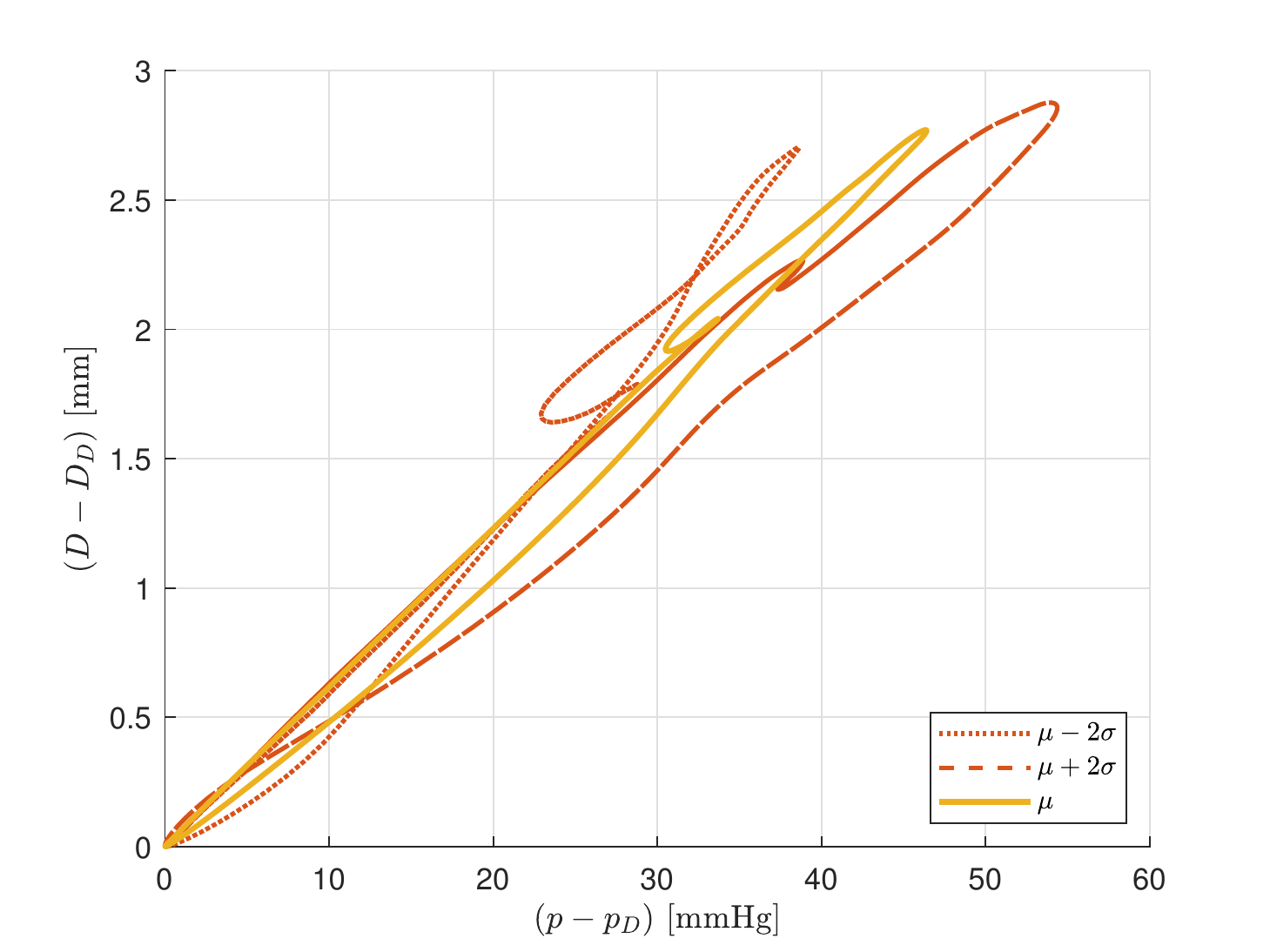}
\vspace*{-5mm}
\caption{}
\label{fig.TC14_visco_hysteresis}
\end{subfigure}
\begin{subfigure}{0.5\textwidth}
\centering
\includegraphics[width=1\linewidth]{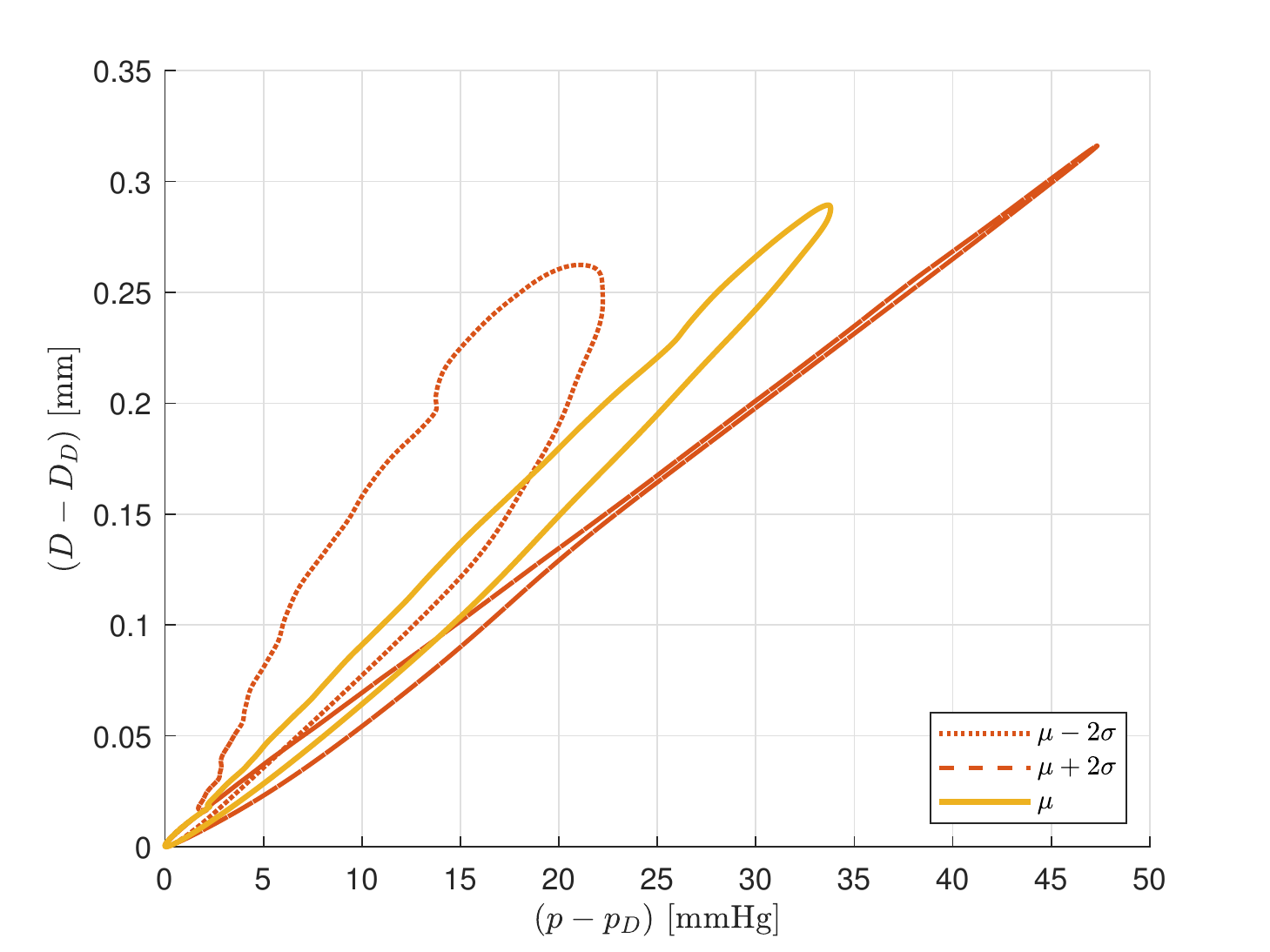}
\vspace*{-5mm}
\caption{}
\label{fig.TC15_visco_hysteresis}
\end{subfigure}
\caption{Numerical hysteresis curves representative of one cardiac cycle obtained in the TA test (a) and CCA test (b) when characterizing the mechanical behavior of the vessel wall through a viscoelastic law. Loops develop in time counter-clockwise and are presented in terms of relative pressure and relative diameter with respect to diastolic values, $p_D$ and $D_D$ respectively. On each plot, the expected hysteresis ($\mu$) is shown, together with those associated to the extreme values of a 95\% confidence interval of the multivariate analysis, evaluated with respect to the standard deviation of both area (diameter) and pressure ($\mu \pm 2\sigma$).}
\label{fig.TC14-15_hysteresis}
\end{figure}
\begin{figure}[t!]
\begin{subfigure}{0.5\textwidth}
\centering
\includegraphics[width=1\linewidth]{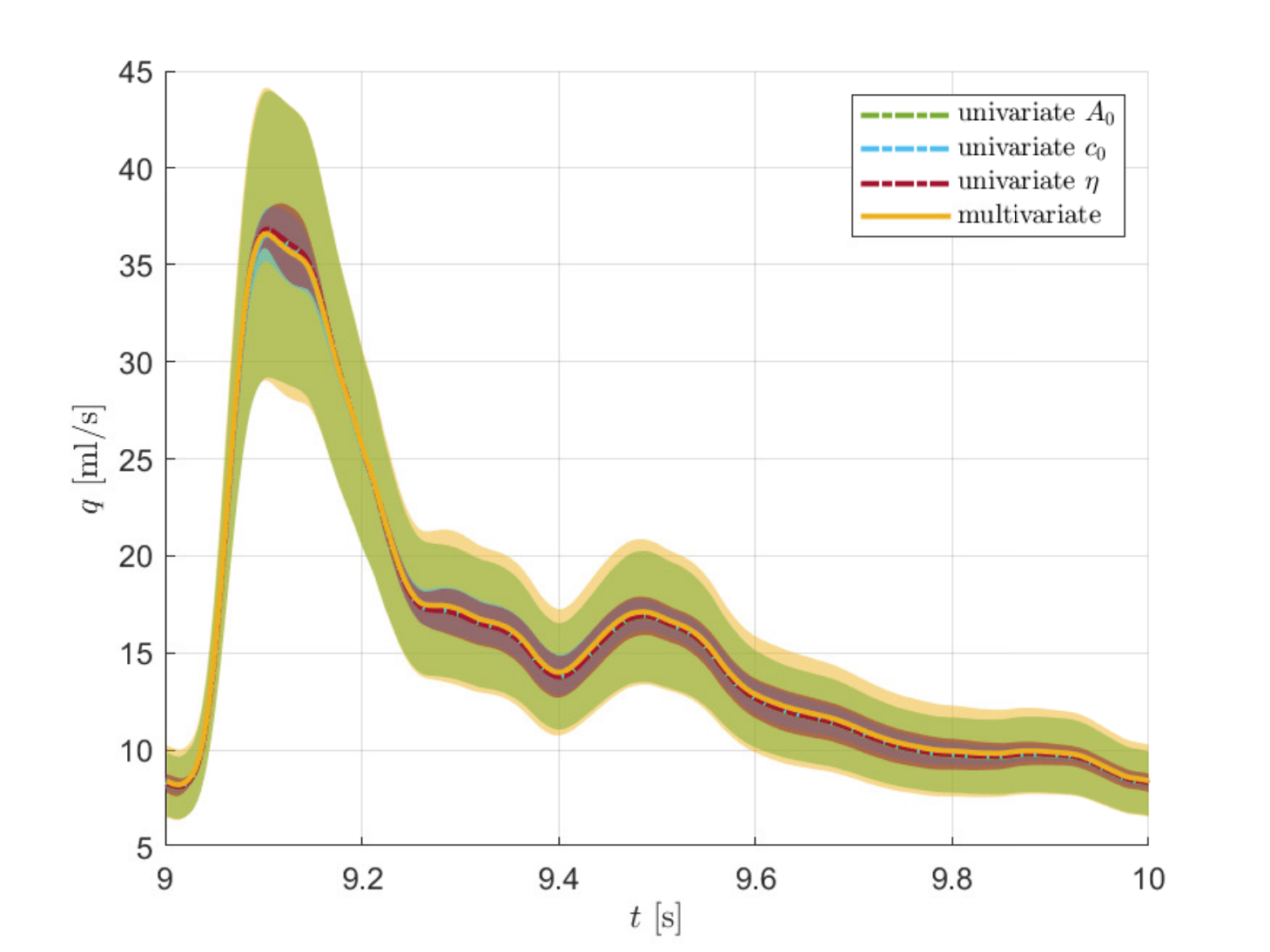}
\vspace*{-5mm}
\caption{}
\label{fig.TC22_visco_q}
\end{subfigure}
\begin{subfigure}{0.5\textwidth}
\centering
\includegraphics[width=1\linewidth]{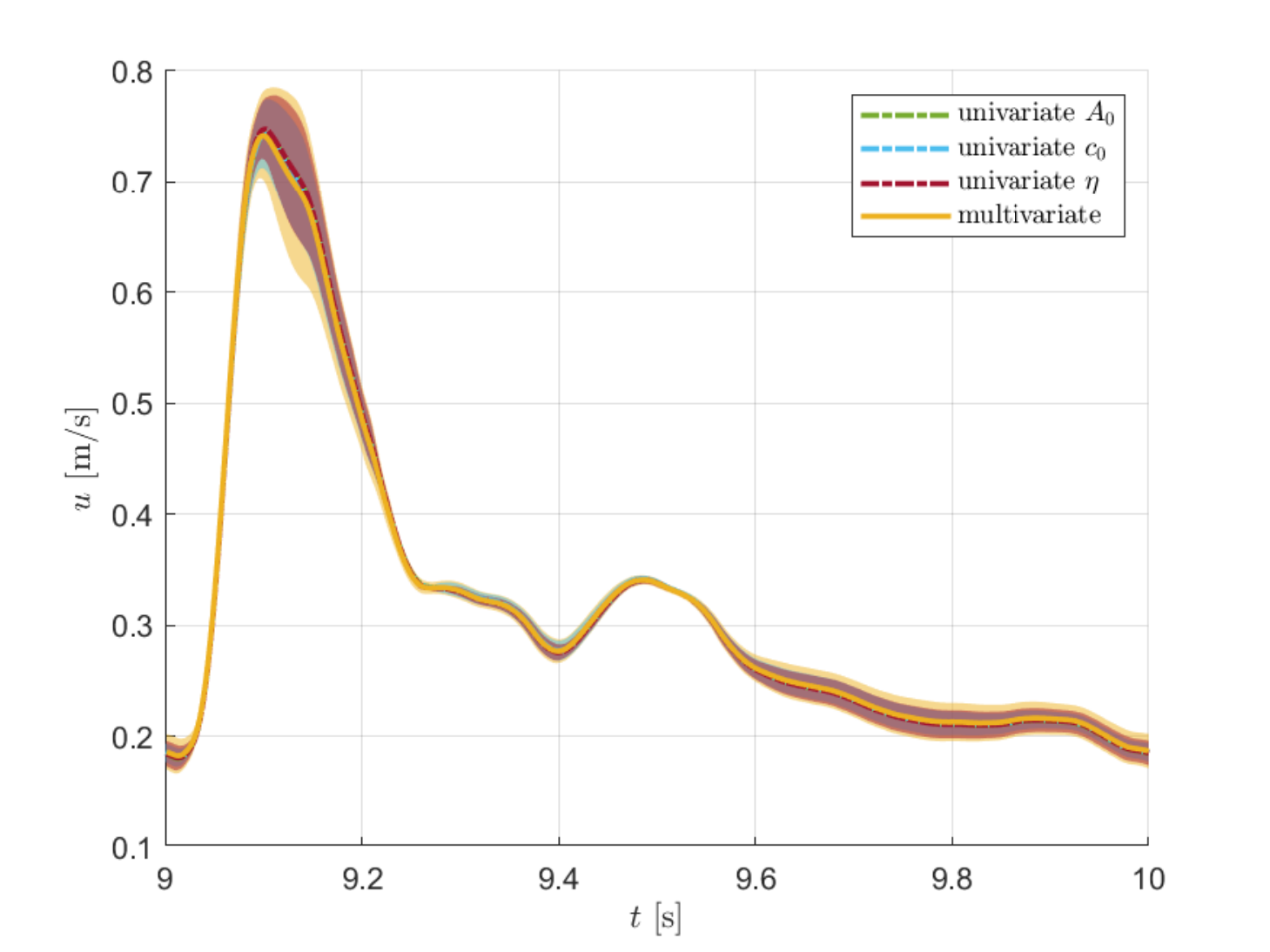}
\vspace*{-5mm}
\caption{}
\label{fig.TC22_visco_u}
\end{subfigure}
\begin{subfigure}{0.5\textwidth}
\centering
\includegraphics[width=1\linewidth]{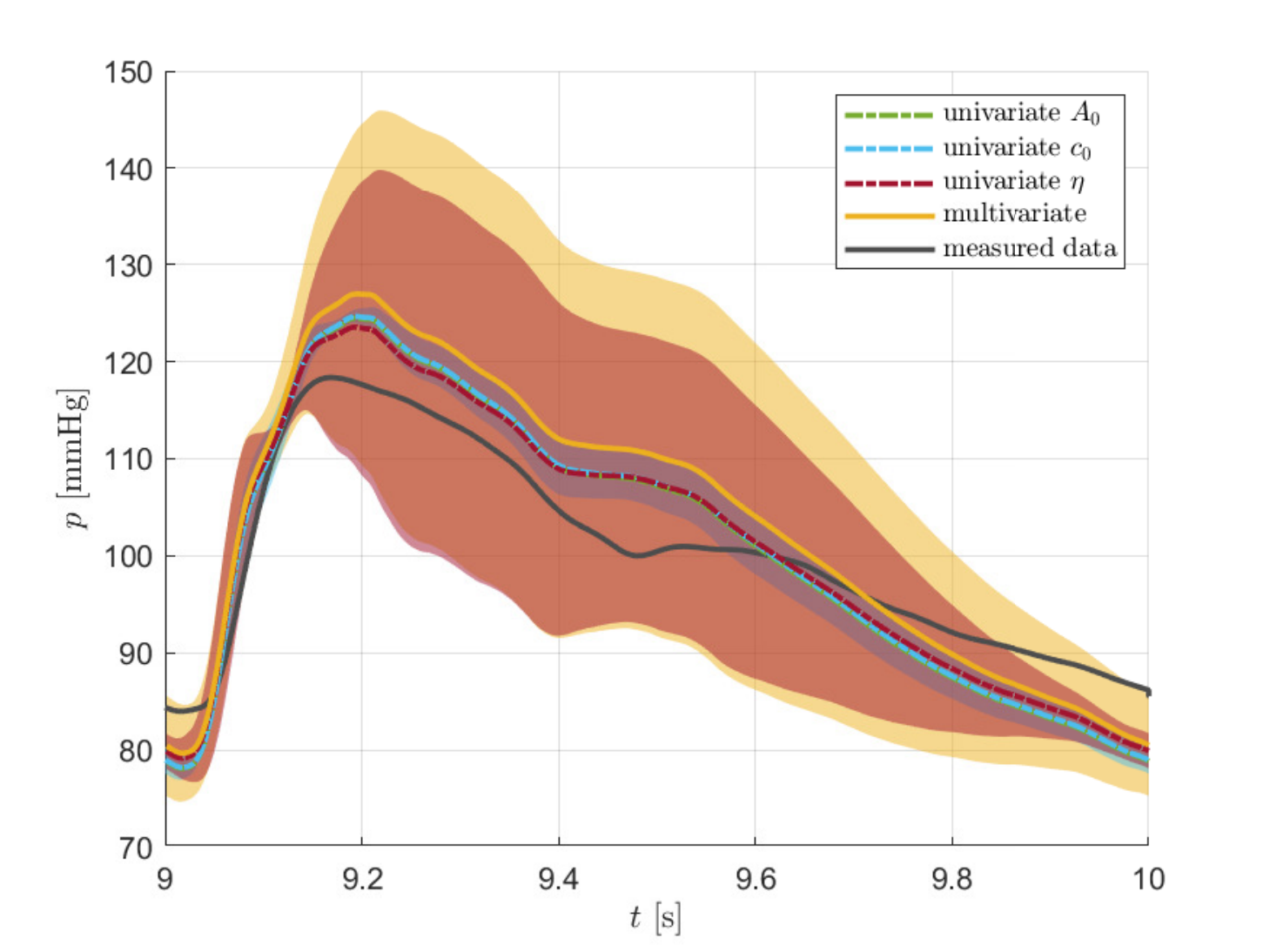}
\vspace*{-5mm}
\caption{}
\label{fig.TC22_visco_p}
\end{subfigure}
\begin{subfigure}{0.5\textwidth}
\centering
\includegraphics[width=1\linewidth]{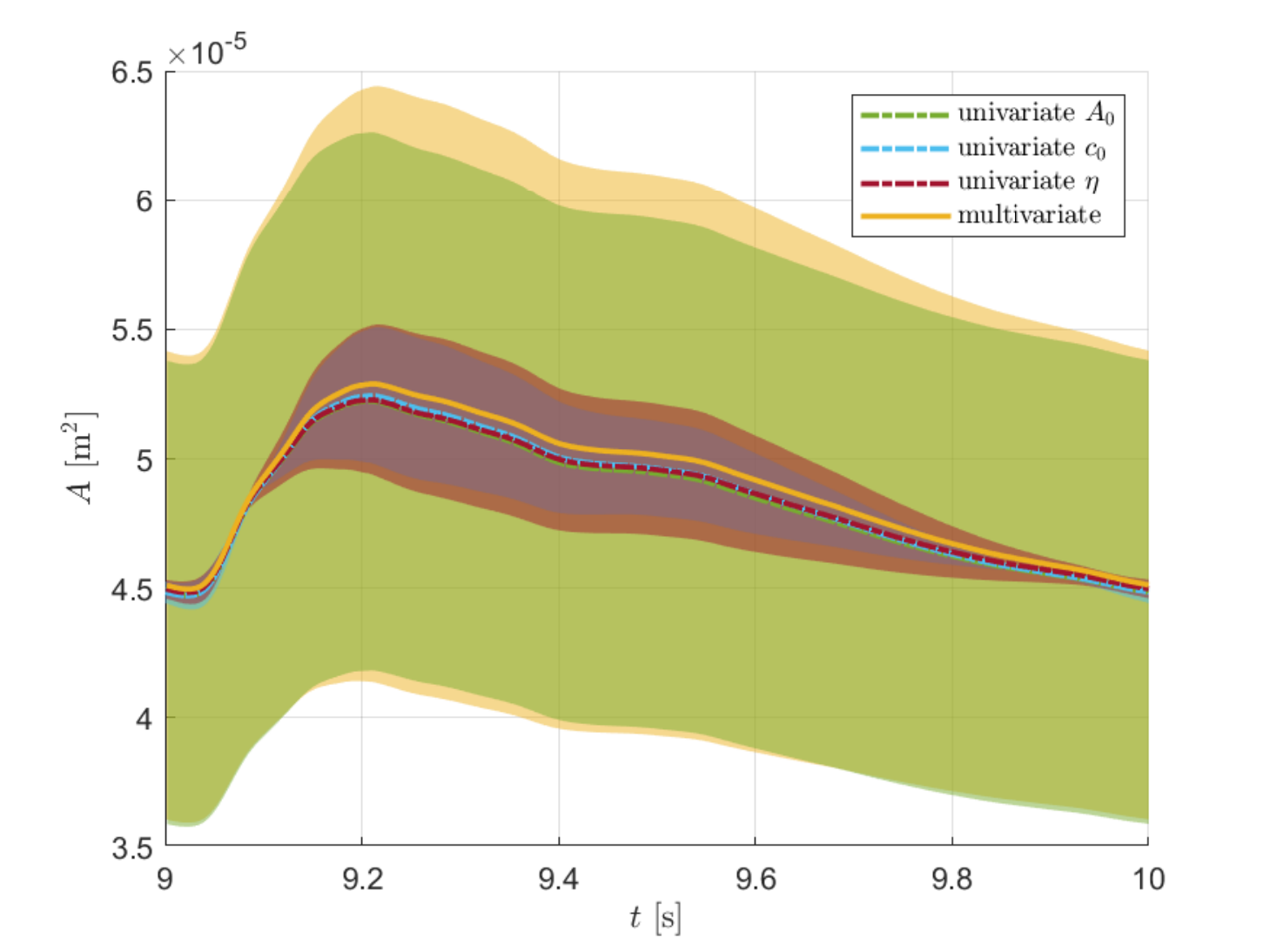}
\vspace*{-5mm}
\caption{}
\label{fig.TC22_visco_A}
\end{subfigure}
\caption{Numerical results representative of one cardiac cycle of patient-specific CCA test (CCA-A), presented in terms of flow rate (a), velocity (b), pressure (c) and area (d) at the midpoint of the domain, for 95\% confidence intervals (colored area) and corresponding expectations (colored line). Each color is associated to a specific simulation, concerning the 3 univariate and the multivariate analysis. Computed pressures are compared to the patient-specific waveform measured in-vivo with the PulsePen tonometer, as described in \cite{bertaglia2020a}.}
\label{fig.TC22_visco}
\end{figure}
\begin{figure}[t!]
\begin{subfigure}{0.5\textwidth}
\centering
\includegraphics[width=1\linewidth]{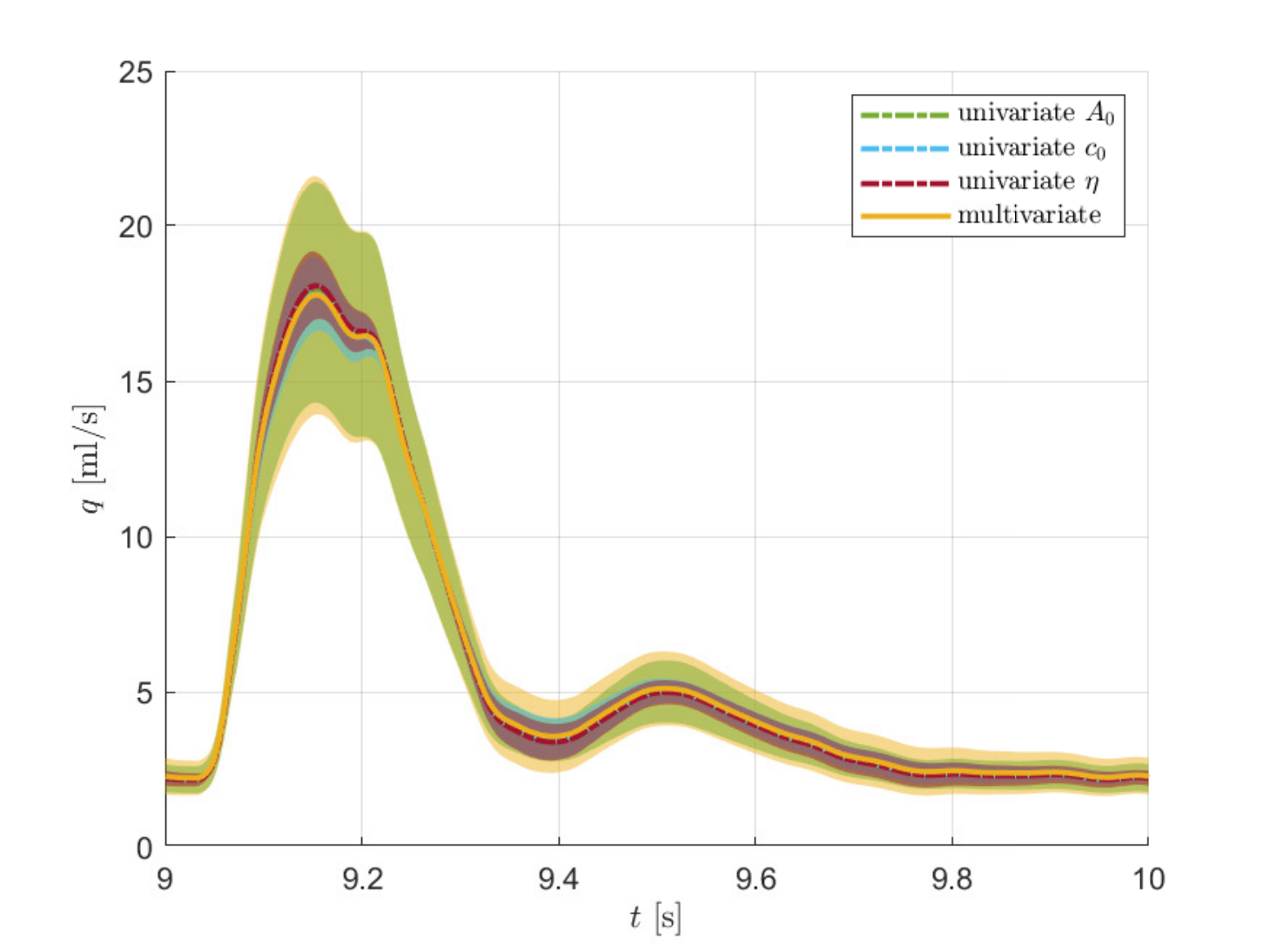}
\vspace*{-5mm}
\caption{}
\label{fig.TC27_visco_q}
\end{subfigure}
\begin{subfigure}{0.5\textwidth}
\centering
\includegraphics[width=1\linewidth]{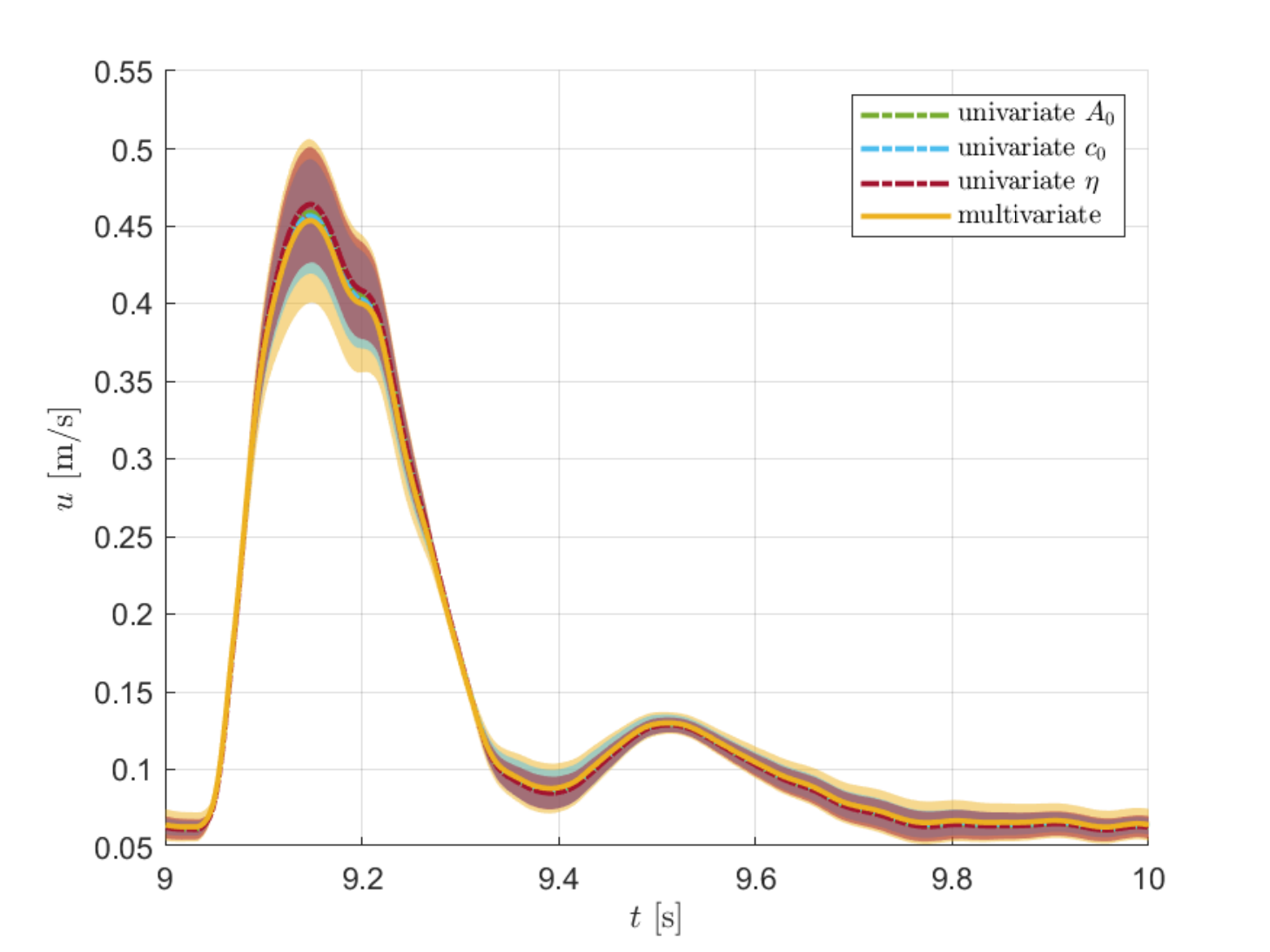}
\vspace*{-5mm}
\caption{}
\label{fig.TC27_visco_u}
\end{subfigure}
\begin{subfigure}{0.5\textwidth}
\centering
\includegraphics[width=1\linewidth]{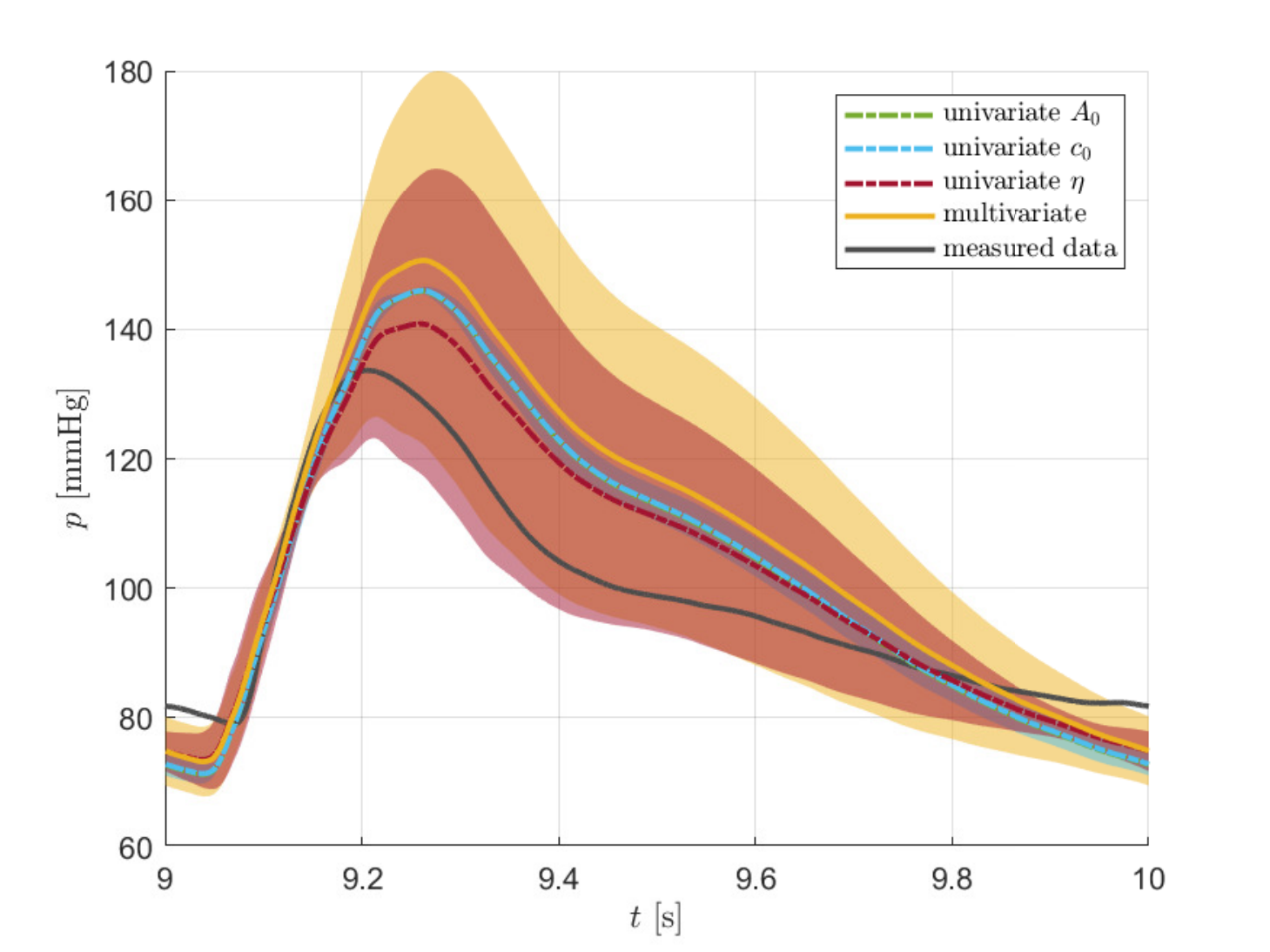}
\vspace*{-5mm}
\caption{}
\label{fig.TC27_visco_p}
\end{subfigure}
\begin{subfigure}{0.5\textwidth}
\centering
\includegraphics[width=1\linewidth]{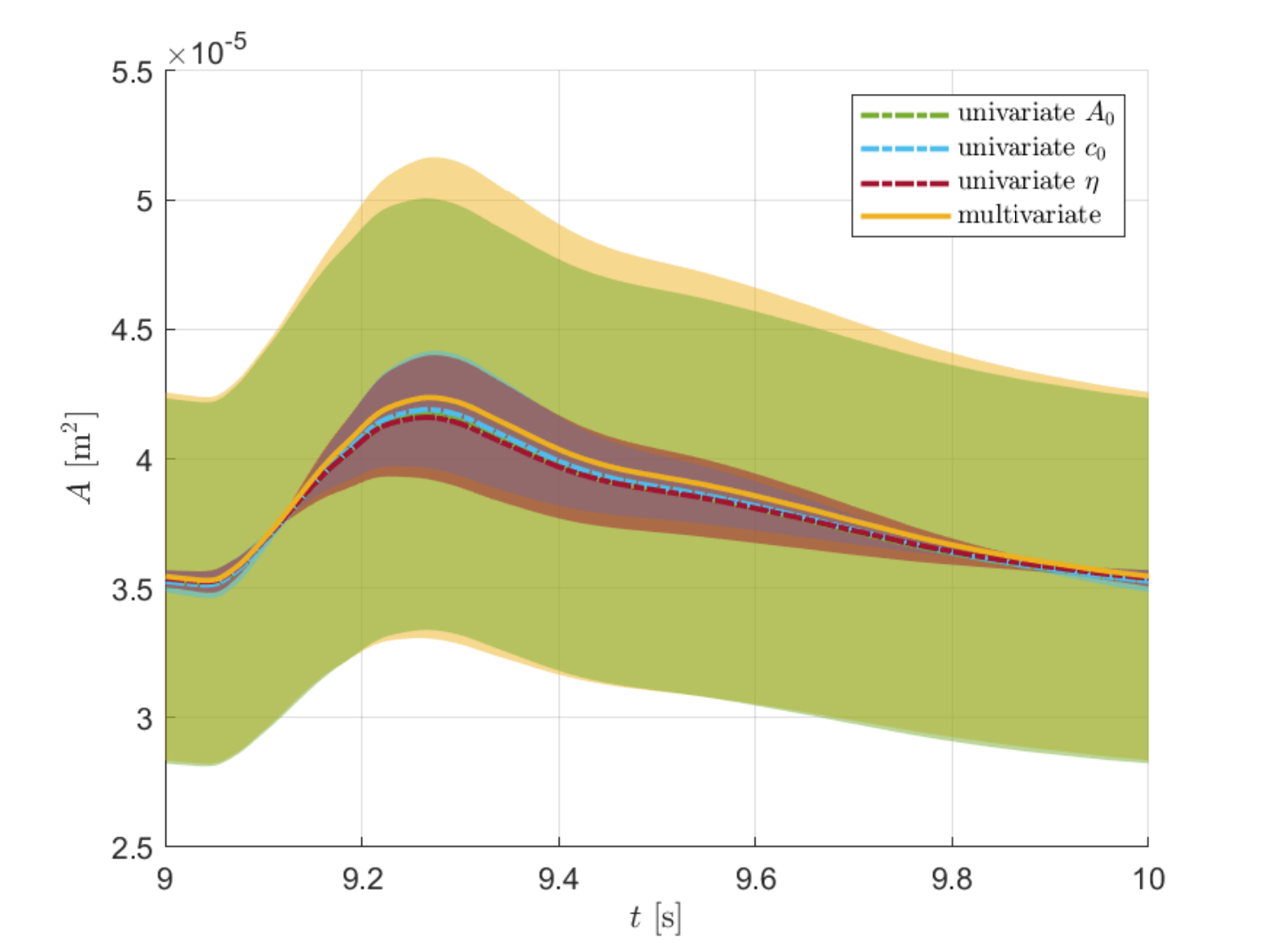}
\vspace*{-5mm}
\caption{}
\label{fig.TC27_visco_A}
\end{subfigure}
\caption{Numerical results representative of one cardiac cycle of patient-specific CFA test (CFA-F), presented in terms of flow rate (a), velocity (b), pressure (c) and area (d) at the midpoint of the domain, for 95\% confidence intervals (colored area) and corresponding expectations (colored line). Each color is associated to a specific simulation, concerning the 3 univariate and the multivariate analysis. Computed pressures are compared to the patient-specific waveform measured in-vivo with the PulsePen tonometer, as described in \cite{bertaglia2020a}.}
\label{fig.TC27_visco}
\end{figure}
\begin{figure}[t!]
\begin{subfigure}{0.5\textwidth}
\centering
\includegraphics[width=1\linewidth]{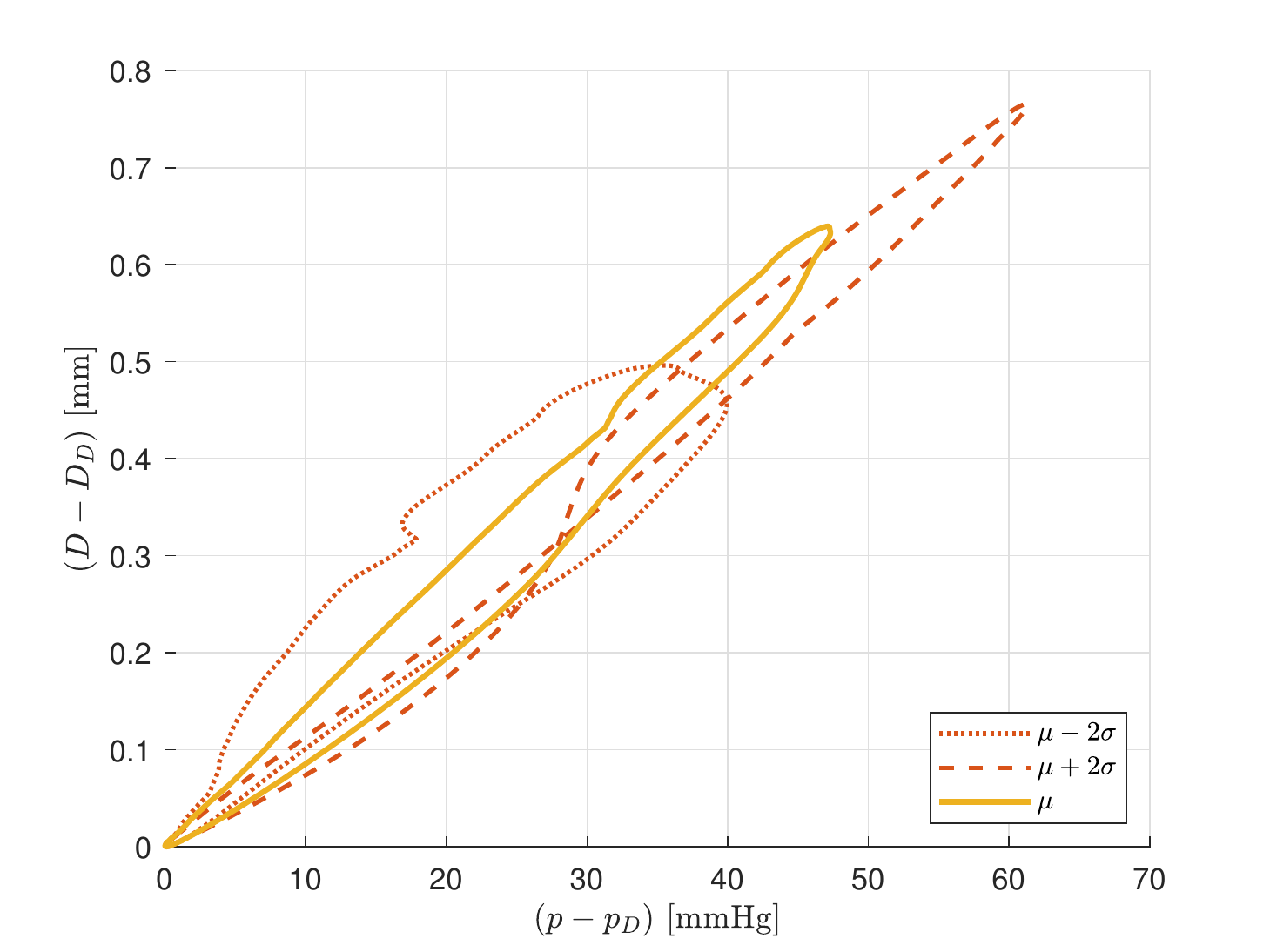}
\vspace*{-5mm}
\caption{}
\label{fig.TC22_visco_hysteresis}
\end{subfigure}
\begin{subfigure}{0.5\textwidth}
\centering
\includegraphics[width=1\linewidth]{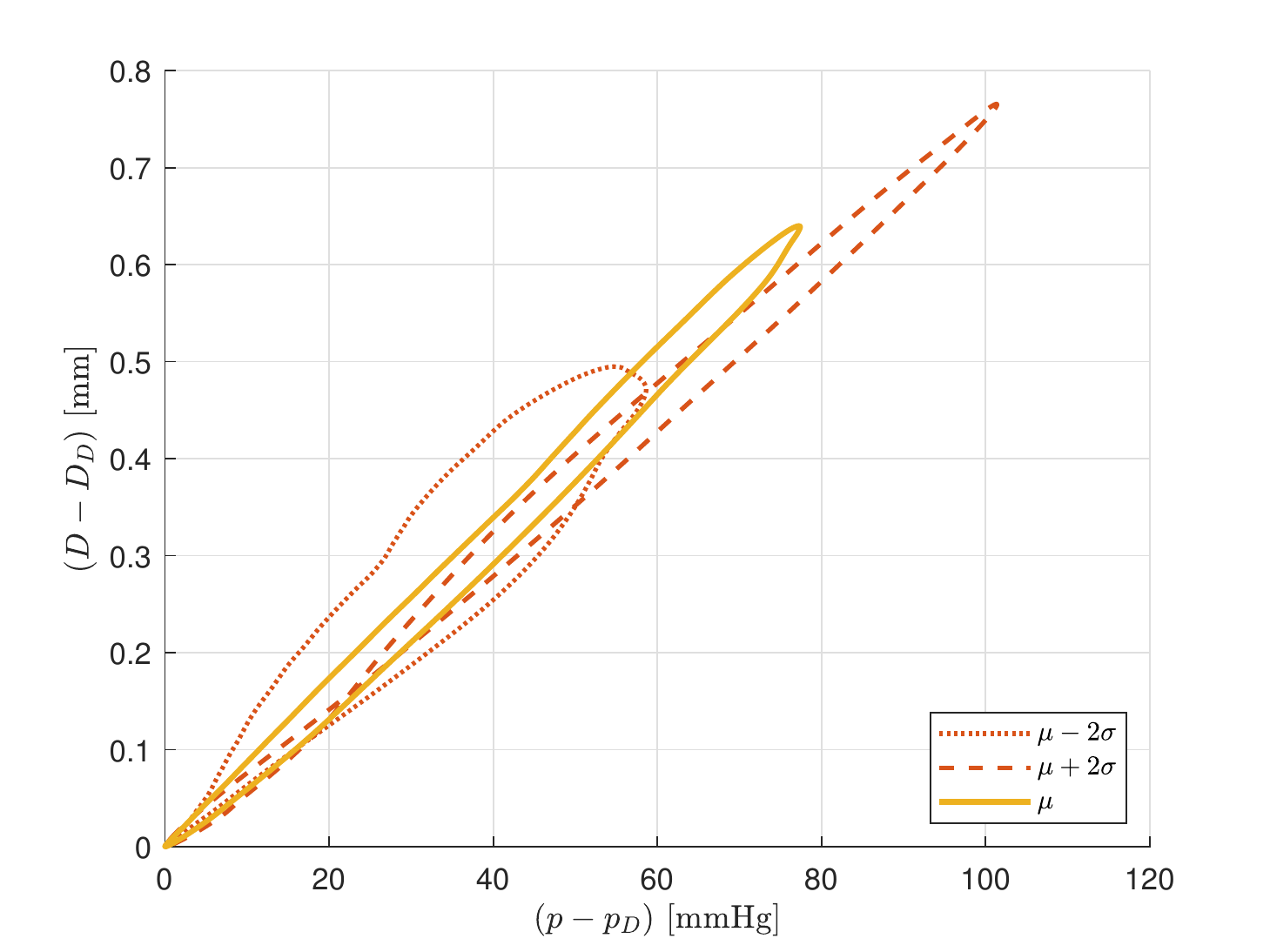}
\vspace*{-5mm}
\caption{}
\label{fig.TC27_visco_hysteresis}
\end{subfigure}
\caption{Numerical hysteresis curves representative of one cardiac cycle obtained in the CCA-A test (a) and CFA-F test (b). Loops develop in time counter-clockwise and are presented in terms of relative pressure and relative diameter with respect to diastolic values, $p_D$ and $D_D$ respectively. On each plot, the expected hysteresis ($\mu$) is shown, together with those associated to the extreme values of a 95\% confidence interval of the multivariate analysis, evaluated with respect to the standard deviation of both area (diameter) and pressure ($\mu \pm 2\sigma$).}
\label{fig.TC22-27_hysteresis}
\end{figure}
\section{UQ applied to the a-FSI blood flow model}
\label{section_UQBloodFlow}
One of the biggest challenges in the clinical application of mathematical models is the adaptation of input data to patient-specific conditions, hence the personalization of models. The inputs to be personalized are included in various categories \cite{arnold2017,chen2013,eck2016}. All measurements of these inputs are hampered by uncertainty as well as large biological variability, leading to uncertainties in the inputs. It has to be considered that very often measurements are not repeated enough times to return reliable statistical estimates. Moreover, not all the inputs can be measured, directly or indirectly (such as the mechanical viscosity of the vessels), and if they are, their measurement may be very expensive, like with the MRI option \cite{eck2016}. In this case, traditional approaches assign the most likely value to these parameters, through parameter estimation procedures, but still leaving a consistent inadequacy to the model, since the behavior of complex biological systems can be very sensitive to these parameters \cite{XIU2007}. Thus, assuming parametric uncertainty, one has then to quantify their impact on the computational results. \par
In this work, we are interested in investigating the effects of uncertainty regarding the parameters involved in the viscoelastic constitutive equation, on which the a-FSI system \eqref{completesyst} is based. Therefore, we assume uncertainty in the equilibrium area $A_0 = A_0(\omega)$, in the reference celerity $c_0 = c_0(\omega)$ and in the viscosity coefficient of the wall $\eta = \eta(\omega)$. With this choice, we are capturing all the uncertainty enclosed in the 3 parameters characterizing the viscoelastic SLS model, estimated in the a-FSI blood flow model as follows \cite{bertaglia2020a}:
\begin{equation}
\label{eq.mech_par}
E_{\infty}(\omega) = \frac{2\rho c_0(\omega)^2 \sqrt{A_0(\omega)}}{h_0 \sqrt{\pi}}, \qquad 
E_0(\omega) = E_{\infty}(\omega) e^{1.3 \cdot 10^{-5}\eta(\omega)}, \qquad 
\tau_r(\omega) = \frac{\eta(\omega) \left(E_0(\omega) - E_{\infty}(\omega)\right)}{E_0(\omega)^2} .
\end{equation}
It is worth to underline that, with the choice of $c_0, A_0$ and $\eta$ as stochastic variables, the model is already accounting for the maximum number of independent random mechanical parameters. In fact, considering as stochastic also the Young modulus would lead to violate the condition of independence of stochastic variables that is at the basis of multivariate analyses \cite{xiu2010,jin2017}, since in the proposed model $E_{\infty} = E_{\infty}(A_0, c_0)$ and $E_0 = E_0(E_{\infty},\eta) = E_0(A_0, c_0,\eta)$. Even though acting in a similar manner on the solution of the problem, $c_0$ and $A_0$ are independent parameters in the proposed methodology. Finally, the uncertainty related to the thickness of the wall $h_0$ and blood density $\rho$ are not taken into account because it has already been demonstrated the minimal impact they have on the solution in central arteries \cite{chen2013}.\par
Generally, for a patient-specific simulation, $A_0$ can be defined recurring to MRI or Doppler ultrasound measurements, the latter being often preferred because of the lower cost, even though possible sources of error and non-optimal accuracy of this technique are well-known \cite{gill1985,park2012}. The carotid-femoral pulse wave velocity (cf-PWV) is considered the gold standard method for the determination of the arterial stiffness (hence $c_0$), which is assessed by dividing traveled distance by travel time \cite{avolio2018}. However, many different procedures have been proposed to determine the value of the carotid-femoral distance, resulting in different cf-PWV values and increasing confusion among users \cite{vanbortel2012}. Finally, there are no effective methods for the measurement of $\eta$. The evaluation of this parameter can be usually performed recurring to hysteresis curves, through mathematical approaches, or, very rarely, carrying ex-vivo tests \cite{armentano1998,kawano2013}.\par
In the absence of sufficient observation for the calibration of the PDF associated to the uncertain parameters, in this work it is assumed a general Gaussian distribution, which reflects errors related to measurements \cite{eck2016}. Furthermore, we consider that deterministic values available in literature and previously used in \cite{bertaglia2020a} are the most likely values, close to the expectation, and that the uncertainty rate does not change the sign of the parameter value. Given the different sources of error related to the above mentioned estimate procedures, different degrees of uncertainty are associated to the three parameters of interest: a medium error of the order of 10\% for $A_0$ and $c_0$ and a 50\% error for $\eta$, which is affected by a more significant uncertainty.\par
To test the impact of these parametric uncertainties on the numerical solution of the a-FSI blood flow model here proposed on single vessel applications, for each test case, three univariate analysis are performed (considering, in turn, only one parameter as stochastic, while keeping the others as deterministic), followed by a multivariate analysis (considering all the three parameters as stochastic). In the following, all the graphical results presented refer to a 95\% confidence interval of the predicted variable of interest (in turn, flow rate, velocity, pressure and area) resulting from each one of the stochastic simulations (2 elastic or 3 viscoelastic univariate stochastic output and one multivariate stochastic output).
\subsection{Numerical Results}
\label{section_resultsBloodFLow}
Referring to deterministic numerical tests presented in \cite{bertaglia2020a}, two initial stochastic simulations are performed taking into account an upper TA and a CCA, with constant-radius over length (baseline arteries). For the characterization of the mechanical behavior of the vessel wall, both the elastic and the viscoelastic constitutive law are considered for these test cases, to evaluate the relevance of the damping effect in terms of sensitivity of the model. Furthermore, two patient-specific tests are evaluated, concerning again a tapered CCA and a tapered CFA, which correspond to tests CCA-A and CFA-F discussed in \cite{bertaglia2020a}, respectively. In tests CCA-A and CFA-F the tapering has been introduced considering a linear variation of the radius $R_0$, while the thickness of the vessel wall $h_0$ is kept constant along the length.\par
Model parameters of each test are here reported in Table~\ref{tab_testdata} for viscoelastic simulations. In the elastic case considered for tests TA and CCA, $E_0 = E_{\infty}$, $\eta \rightarrow 0$ and $\tau_r \rightarrow 0$, with the resulting source term $S=0$, as discussed in \S~\ref{section_IMEX}. Concerning numerical parameters, $\CFL = 0.9$ in all the simulations. The number of cells in the physical domains is $N_x = 12$ in the TA test, $N_x = 6$ in the CCA test and $N_x = 7$ in the two patient-specific cases. It is here underlined that, with respect to the values of $\tau_r$ reported in Table~\ref{tab_testdata}, the chosen discretization leads to deal with weakly stiff problems in the viscoelastic case. For each stochastic parameter investigated, $A_0, c_0$ and $\eta$ (the latter considered only when concerning the viscoelastic tube law), $N_p = 3$ collocation points resulted enough in the light of what discussed in \S~\ref{section_resultsBurgers}, with a total of 27 collocation points involved in the viscoelastic multivariate analysis (9 collocation points in the elastic multivariate analysis). In this configuration, indeed, the contribution of the stochastic mesh size results comparable to the physical mesh size one and further refinement of the stochastic grid, augmenting the number of collocation points, does not increase the overall accuracy of numerical solutions \cite{petrella2019a}. For the TA case, 20 cardiac cycles are simulated, corresponding to a final time $t_{end} = 19.10$ s; while for the CCA case, 9 cardiac cycles are simulated, corresponding to a final time $t_{end} = 9.90$ s. In both the patient-specific simulations, 10 cardiac cycles are reproduced, corresponding to a final time $t_{end} = 10.00$ s.
\subsubsection{Baseline arteries}
\label{section_testsBaseline}
Numerical results of the stochastic problems concerning baseline TA and CCA characterized by a simple elastic behavior of the wall are shown in Figs.~\ref{fig.TC14_elastic} and \ref{fig.TC15_elastic}, respectively. For each test case, results of all the univariate and multivariate analysis are shown together in a single plot, to better highlight differences in terms of expected waveforms and confidence intervals.\par
First of all, it is clearly visible that expected waveforms result very similar in each type of stochastic analysis, being even indistinguishable in certain cases. Concerning confidence intervals, in Fig.~\ref{fig.TC14_elastic} it can be observed that flow rate, and in part also velocity waveforms, are very little sensitive to all the parametric uncertainties investigated in the elastic TA case. On the other hand, for pressure and area (which evolve in synchrony in the elastic case) it is sufficient a moderate change (10\% error) of geometric and mechanical characteristics to cause large variabilities in the final solution. As expected, the uncertainty on $c_0$ has a major impact on the pressure trend, while uncertainty on $A_0$ mostly affects area and velocity variation over time. It can be verified that these results are in line with literature findings \cite{chen2013,petrella2019a}. 
Similar observations can be made concerning results of the CCA case, shown in Fig.~\ref{fig.TC15_elastic}. Here, a higher sensitivity of the velocity waveform is reported with respect to the uncertainty related to $A_0$ and in the joint effect produced by both the two stochastic parameters (Fig.~\ref{fig.TC15_elastic_u}). This more visible sensitivity is probably due to the reduced magnitude of the velocity in the CCA with respect to the TA. \par
Results obtained taking into account a more realistic viscoelastic FSI characterization are presented in Figs.~\ref{fig.TC14_visco} and \ref{fig.TC15_visco}, with corresponding qualitative hysteresis curves shown in Fig.~\ref{fig.TC14-15_hysteresis}. Again, a considerable difference in the sensitivity of flow rate and velocity with respect to the sensitivity of pressure and area emerges, with the same characteristics discussed for the elastic case. 
However, it can be noticed that due to the very high uncertainty (50\%) associated to the viscosity parameter $\eta$, simulation outputs show that in pressure confidence intervals a prominent role is played by the wall viscosity. Nevertheless, the uncertainty of $\eta$ (by which pressure is most affected) principally impacts the systolic peak and the dicrotic limb, not acting on the anacrotic limb.\par
Finally, in Fig.~\ref{fig.TC14-15_hysteresis}, hysteresis loops of multivariate tests TA and CCA generated with the expected values of $p$ and $A$ (expressed in terms of diameter $D$ in the plot), which are evaluated with \eqref{eq:mean_apx}, are shown together with upper and lower bounds of the 95\% confidence interval obtained adding and subtracting twice the corresponding standard deviation, which is the square root of the variance computed with \eqref{eq:variance_apx}, at each of the two variables' mean prediction. It can be observed that hysteresis curves resulting from the multivariate analysis vary in response of the parametric uncertainties both in slope and amplitude, since all the three viscoelastic parameters of the SLS model are involved by the stochastic characterization \cite{bertaglia2020a}.
\subsubsection{Patient-specific arteries}
\label{section_testsInvivo}
Numerical results of the stochastic problems concerning patient-specific CCA and CFA are presented in Figs.~\ref{fig.TC22_visco} and \ref{fig.TC27_visco}, respectively, with the corresponding qualitative hysteresis curves shown in Fig.~\ref{fig.TC22-27_hysteresis}. As for the previous tests, for each artery, results of all the univariate and multivariate analysis are shown together in a single plot, to permit immediate comparisons. The sensitivity emerging from the patient-specific cases well reflects the one registered in the baseline tests. Indeed, similar differences in the variability of flow rate and velocity emerge with respect to the variability of pressure and area, the latter resulting more sensitive to the parametric uncertainties underlying geometrical and mechanical characterization of vessels. However, a slightly higher sensitivity in the flow rate waveforms can be noticed in Figs.~\ref{fig.TC22_visco_q} and \ref{fig.TC27_visco_q} with respect to Figs.~\ref{fig.TC14_visco_q} and \ref{fig.TC15_visco_q}. This difference reflects the diverse inlet boundary condition used in the two type of tests: while in the baseline tests it has been imposed an inlet flow rate available in literature \cite{boileau2015}, in CCA-A and CFA-F tests only a velocity waveform extrapolated from patient-specific Doppler measurements could be used as inlet condition \cite{bertaglia2020a}. Hence, in the patient-specific tests, the flow rate carries the uncertainty associated to $A_0$.\par
In Figs.~\ref{fig.TC22_visco_p} and \ref{fig.TC27_visco_p}, expectations of the computed pressures are compared to patient-specific waveforms measured in-vivo, respectively on the CCA and on the CFA, by means of the arterial applanation tonometry technique (the reader can refer to \cite{bertaglia2020a} for details on data acquisition and extrapolations). Concerning both the types of artery, confidence intervals resulting from multivariate analysis well capture the in-vivo signal. Expected pressure waveforms are comparable to the measured ones, especially in the CCA case, with a little overestimation of the systolic value in the CFA case. Finally, by looking at hysteresis loops presented in Fig.~\ref{fig.TC22-27_hysteresis}, we can affirm that when geometrical and mechanical parameters involved in the constitutive law governing the FSI mechanism are underestimated with respect to the expected value, hysteresis curves appear enlarged in amplitude; vice-versa, when these parameters are overestimated, loops get narrower.
\section{Conclusions}
\label{section_conclusions}
In the present work, the effects of uncertainties of the parameters involved in the elastic and viscoelastic characterization of the FSI dynamics occurring between arterial wall and blood flow have been investigated through univariate and multivariate UQ analysis based on an IMEX finite volume stochastic collocation approach. The proposed method combines a finite volume solver suitable for the non-conservative terms of the hyperbolic model with an IMEX method that satisfies the AP property for small relaxation times.\par 
The methodology has been validated using a model equation through a thorough convergence study, which confirms the spectral accuracy of the stochastic collocation method and the second-order accuracy of the chosen IMEX Runge-Kutta finite volume method. 

Concerning UQ analyses applied to the a-FSI blood flow model, a different sensitivity emerges when comparing the variability of flow rate and velocity waveforms to the variability of pressure and area, the latter ones resulting much more sensitive to the parametric uncertainties considered. 
When comparing variables predicted adopting the elastic tube law with respect to the viscoelastic one, it can be noticed that, as expected, the great uncertainty of the viscosity parameter $\eta$ ($\pm 50\%$ of possible error) plays a major role in the output of pressure waveforms, enlarging the confidence interval of this variable. Nevertheless, the impact of the wall viscosity is principally visible in the systolic peak and dicrotic limb and not in the anacrotic limb. 
Moreover, at the clinical level, results suggest that, with respect to the same uncertainty of the parameters taken into account in the proposed work, there is greater uncertainty in predicting a patient's accurate area waveform than predicting a pressure waveform. In fact, in the latter, the chosen stochastic parameters impact the systolic peak and the dicrotic limb, leaving basically untouched the anacrotic limb.
Concerning hysteresis curves, we have observed how the underestimation of geometrical and mechanical parameters, involved in the viscoelastic constitutive law, gives rise to enlarged hysteresis loops; while the overestimation of them results in narrower loops.
Finally, in patient-specific tests, confidence intervals resulting from multivariate analysis well capture in-vivo measured data and expectations of the computed pressures are comparable to recorded waveforms, confirming that the proposed model is a valuable tool for improving cardiovascular diagnostics.
 
Future perspectives include UQ analyses of the stochastic a-FSI blood flow model concerning arterial networks (including proper junction conditions), to assess the effects of elastic and viscoelastic parameters in more complex domains. With this intent, sparse grids techniques, aimed at increasing the efficiency of the method even for simulations with a very large number of random inputs (like those involved in extended networks), will be investigated.
\section*{Acknowledgements}
This work was partially supported by MIUR (Ministero dell'Istruzione, dell'Universit\`a e della Ricerca) PRIN 2017, project \textit{``Innovative numerical methods for evolutionary partial differential equations and applications''}, code 2017KKJP4X. The second author, Valerio Caleffi, was also funded by MIUR FFABR 2017.
\appendix
\section{Quasi-exact solution of the vBE with Gaussian initial conditions}
\label{appendix:A}
The IVP \eqref{eq:IV_problem} can be solved applying the Cole–Hopf transformation \cite{Cole1951,Hopf1950}:
\begin{equation}
\label{eq:CH_trasf}
q(x,t) = - 2 \nu \frac{\partial_x \varphi(x,t)}{\varphi(x,t)},
\end{equation}
to obtain the new simpler IVP based on the diffusion equation in the dependent variable $\varphi(x,t)$:
\begin{subequations}
\label{eq:IV_diff_problem}
\begin{align}
&\partial_t \varphi(x,t) = \nu \partial_{xx} \varphi(x,t); \label{eq:diff_det} \\
&\varphi(x,0) = \varphi_0(x); \label{eq:diff_IV}
\end{align}
\end{subequations}
which solution reads:
\begin{equation}
\label{eq:diff_sol_gen}
\varphi(x,t) = \int_{-\infty}^{+\infty} \frac{\varphi_0(\xi)}{\sqrt{4\pi\nu t}}\exp\left[-\frac{(x-\xi)^2}{4\nu t} \right] \d \xi.
\end{equation}
To write an explicit quasi-exact solution, let us consider the Gaussian initial condition presented in Eq.~\eqref{eq:vBE_ic_gs}. Using the Cole–Hopf transformation \eqref{eq:CH_trasf}, after an integration by parts, the initial condition can be expressed in terms of $\varphi(x,t)$, in the form:
\begin{equation}\label{eq:diff_ic_gs}
\varphi(x,0) = \varphi_0(x) = \exp\left[ - \frac{\sigma_m q_{0,m}}{2 \nu}
\sqrt{\frac{\pi}{2}} \erf\left( \frac{x}{\sqrt{2}\sigma_m}\right) \right] .
\end{equation}
The solution of the IVP \eqref{eq:diff_det}-\eqref{eq:diff_ic_gs} is formally given by Eq.~\eqref{eq:diff_sol_gen} with $\varphi_0(x)$ given by Eq.~\eqref{eq:diff_ic_gs}. For the specific initial conditions considered here, Eq.~\eqref{eq:diff_sol_gen} cannot be analytically solved and a numerical procedure is needed.\par
A simple approach consists in the introduction of the auxiliary variable:
\begin{equation}
\eta = \frac{x-\xi}{\sqrt{4 \nu t}},
\end{equation}
and in the selection of a Gauss-Hermite quadrature \cite{Abramowitz_Stegun1972}, which allows to express the solution \eqref{eq:diff_sol_gen} as:
\begin{equation}\label{eq:diff_sol_GH}
\varphi(x,t) = -\frac{1}{\sqrt{\pi}}\int_{-\infty}^{+\infty} \varphi_0(x - \eta\sqrt{4\nu t})\,e^{-\eta^2} \d \eta \approx -\frac{1}{\sqrt{\pi}} \sum_{j=1}^{n_{GH}} \varphi_0(x - \eta_j \sqrt{4\nu t})\, w_j,
\end{equation}
where $\eta_j$ and $w_j$ are nodes and weights of the quadrature.\\
Similarly, the space derivative of $\varphi$ can be expressed as:
\begin{equation}\label{eq:diff_dx_GH}
\partial_x \varphi(x,t) = \frac{1}{\sqrt{\pi}}\int_{-\infty}^{+\infty} \varphi_0(x - \eta\sqrt{4\nu t})\,\frac{\eta}{\sqrt{\nu t}}\,e^{-\eta^2} \d \eta \approx \frac{1}{\sqrt{\nu t}} \frac{1}{\sqrt{\pi}} \sum_{j=1}^{n_{GH}} \eta_j\,\varphi_0(x - \eta_j \sqrt{4\nu t})\, w_j.
\end{equation}
Finally, the approximated solution $q(x,t)$ can be computed using Eq.~\eqref{eq:CH_trasf}.
\section{IMEX Runge-Kutta scheme applied to the vBE}
\label{appendix:B}
For a numerical solution of the vBE based on the application of the second-order IMEX Runge-Kutta finite volume method described in \S~\ref{section_IMEX}, Eq.~\eqref{eq:vBE_det} is written in the conservative form
\begin{equation}\label{eq:vBE_cons}
\partial_t q + \partial_x f(q) = g(q),
\end{equation}
where the analytical flux is $f(q) = \frac{q^2}{2}$ and the diffusive part is treated as a source term $g(q)=\nu\partial_{xx}^2q$. Hence, the IMEX discretization proposed in Eqs.~\eqref{eq.IMEX} becomes:
\begin{subequations}
\begin{align}
	&q^{(k)}_i = q^n_i -  \frac{\Delta t}{\Delta x} \sum_{j=1}^{k-1} \tilde{a}_{kj}
	 \left( \hat{f}_{i + \frac{1}{2}}^{(j)}  - \hat{f}_{i - \frac{1}{2}}^{(j)} \right) + \Delta t \sum_{j=1}^{k} a_{kj}\, \hat{g}_i^{(j)}
	\label{eq.iterIMEXvBE} \\
	& q^{n+1}_i = q^n_i -  \frac{\Delta t}{\Delta x} \sum_{k=1}^{s} \tilde{b}_{k}  \left( \hat{f}_{i + \frac{1}{2}}^{(k)}  - \hat{f}_{i - \frac{1}{2}}^{(k)} \right) + \Delta t \sum_{k=1}^{s} b_{k}\, \hat{g}_i^{(k)}, \label{eq.finalIMEXvBE}
\end{align}\label{eq.IMEXvBE}
\end{subequations}
where $\hat{f}_{i \pm \frac{1}{2}}$ are the numerical fluxes and $\hat{g}_{i}$ is a suitable discretization of the source term. The direct application of the path-conservative DOT solver to Eq.~\eqref{eq:vBE_cons} permits to define the numerical flux as:
\begin{equation}
\label{eq:flux_vBE}
	\hat{f}_{i+\frac{1}{2}} = \frac{1}{2} \left[ f\left(q_{i+\frac{1}{2}}^{+}\right) + f\left(q_{i+\frac{1}{2}}^{-}\right)\right] - \frac{1}{2} \int_{0}^{1} \left \lvert \Psi\left(q_{i+\frac{1}{2}}^{-},q_{i+\frac{1}{2}}^{+},s\right)\right \rvert \frac{\partial \Psi}{\partial s}\d s ,
\end{equation}
where the selection of a linear segment path $\Psi$ is again sufficient to obtain the required accuracy.\\
The diffusive term can be simply discretized by second-oder central finite differences.
The use of this approximation, together with the explicit treatment of the homogeneous part of Eq.~\eqref{eq:vBE_cons}, leads to a simple tridiagonal system of equations, which is solved at each stage of the IMEX Runge-Kutta method using the Thomas algorithm. 
\bibliographystyle{abbrv}
\bibliography{UQBloodFlow_Bertaglia2020}

\end{document}